\documentclass[final]{siamltex}
\pdfoutput=1

\usepackage{graphicx}
\usepackage{epstopdf,cite,url}
\usepackage{wrapfig,colortbl,multirow}
\usepackage{amsmath,amssymb,amsfonts}

\definecolor{kcolor}{rgb}{ 1, 0.65, 0}
\definecolor{methcolor}{rgb}{ 0.3320312, 0.4179688, 0.1835938}

\newcommand{\Hypk}[0]{H_j}
\newcommand{\Hyps}[0]{\mathcal{H}}
    
    \newcommand{\HypsFix}[1]{\mathcal{H}^{fix}_{#1}}
    \newcommand{\HypsHid}[1]{\mathcal{H}^{hid}_{#1}}
    \newcommand{\HypsSample}[0]{\mathcal{H}^{samp}}
      \newcommand{\CSample}[0]{\mathcal{C}^{samp}}
    \newcommand{\ClusterSample}[1]{\mathcal{C}^{samp}_{#1}}

\def\RR{{\color{red}R}} 
\def\CC{{\color{red}C}}

\def\AA{{\color{red}A}} 
\def\BB{{\color{red}B}} 

\newcommand{\twopartdef}[4]
{
    \left\{
        \begin{array}{ll}
            #1 & \mbox{if } #2 \\
            #3 & \mbox{if } #4
        \end{array}
    \right.
}

\newcommand{\twopartdefother}[3]
{
    \left\{
        \begin{array}{ll}
            #1 & \mbox{if } #2 \\
            #3 & \mbox{otherwise}
        \end{array}
    \right.
}

\title{Uncertainty quantification of discrete association problems in image sequence-based  tracking
\thanks{This work was partially supported by Numerica Corporation and a grant from the AFOSR. Mont and Calderon (Numerica Corporation) designed the algorithms described herein and generated all numerical results.  Weigel and Krapf (Colorado State University) conducted quantum dot imaging experiments. Corresponding Author: chris.calderon@numerica.us }}

\author{Alexander Mont  \and Aubrey V. Wiegel  \and Diego Krapf  \thanks{Electrical \& Computer Engineering and School of Biomedical Engineering, Colorado State University, Fort Collins, Colorado, 80523}
        \and \ \ \ \ Christopher P. Calderon \thanks{Numerica Corporation, 4850 Hahns Peak Drive, Loveland, Colorado, 80538 
        }} 

\begin{document}

\maketitle

\begin{abstract} Applications, ranging from tracking molecular motion within cells to analyzing complex animal 
foraging behavior, require algorithms for associating  a collection of spot-like particles in one image 
with particles contained in another image. These associations are often made via  network flow algorithms.
However, it is often the case that many candidate association solutions (the output of network flow algorithms)  
have nearly optimal scores;  in this case,  the optimal assignment solution is of dubious quality.
Algorithms for reliably computing the uncertainty of candidate association solutions are under-developed in 
situations where many particles are tracked over multiple frames of data.  This is due in part to the fact that 
exact uncertainty quantification (UQ) in large association problems  is computationally intractable because 
the exact computation exhibits exponential dependence on the number of particles tracked.   We introduce a 
technique that can accurately and efficiently quantify association ambiguity (i.e., UQ for discrete association 
problems) without requiring the evaluation of the cost of each feasible association solution. Our method  can readily be wrapped around  existing tracking algorithms and can efficiently handle a variety of 2D association  problems. 
The applications presented are focused on tracking molecules in live cells. Our method is validated via both simulations and  experiments. The experimental applications aim to accurately form tracks and quantify diffusion of quantum
dot labeled proteins from \emph{in vivo} measurements; here  association problems involving cost matrices possessing hundreds to thousands of rows/columns are encountered. For such large-scale problems, we discuss how our approach can 
efficiently and accurately quantify inherent uncertainty in candidate data associations.

\end{abstract}



\pagestyle{myheadings}
\thispagestyle{plain}

\section{Introduction}

Tracking multiple point-like objects exhibiting complex (\emph{a priori} unknown) dynamics in a crowded environment
over a sequence of images with the intention of inferring statistical kinetic information
is of interest to a variety of applications.
For example, the work presented herein was inspired by advances in optical microscopy which have significantly increased the temporal and spatial resolution associated with  measurements of particles in live cells \cite{Kim2006,Danuser2011,Meijering2012}.
The ability to  accurately monitor fluorescently tagged molecules and/or cells in their native biological  environment  
has enormous potential in addressing open questions in molecular and cell biology \cite{Saxton2008,Genovesio2006,Kim2006,Manley2008,Serge2008,Jaqaman2008,Meijering2012}.
In these applications,  experiments produce a large volume of time-ordered image data that require extensive computational processing 
before biologically relevant information can be extracted \cite{Genovesio2006,Serge2008,Jaqaman2008}. 
Although the remainder of this work is focused on tracking molecules within cells, the techniques are applicable to a variety of diverse scenarios where the goal is to gather temporal statistics on multiple objects measured simultaneously in images  \cite{Saligrama2010,Cezar2010,Fox2011}.

Track formation is necessary before particle kinetics can be inferred from experimental data. Throughout this article, we use the term ``track"  to refer to a collection of measurements at different times believed to be produced by the same underlying  molecule. 
The quality of tracks formed via association problems  \cite{Kragel2012} 
  can vary substantially due to various factors \cite{Danuser2011,Meijering2012}.  For example, the environment associated with the collected data may be cluttered, hence complicating track formation  \cite{Kim2006,Saxton2008}  or the observed physical motion of the particle may be different than that assumed by models attempting to forecast particle motion.
Therefore, computational means for automatically and efficiently quantifying the uncertainty in candidate tracks 
provides valuable information that can be used for the assessment of track formation.
 State-of-the-art single particle trackers \cite{Genovesio2006,Serge2008,Jaqaman2008,Danuser2011,Meijering2012} in biological applications focus on extracting the optimal association  solution and do not directly quantify  the uncertainty in association solutions.  The uncertainty in   discrete association problems is fundamentally different than track state uncertainty  (which is often approximated via covariance matrices).

 In this article, we introduce a computationally efficient approach for quantifying association uncertainty  (also known as ``ambiguity" \cite{Kragel2012}).  The approach is illustrated on both controlled simulations and experimental data.
 Our algorithm   
 can efficiently quantify  the probability of putative associations 
connecting two distinct frames, i.e., questionable tracks are flagged without requiring  ground-truth data guidance. 
This output can quantify both frame-to-frame and ``gap-closing'' associations \cite{Jaqaman2008} where track segments at different times are stitched together.
In biological tracking, this has great utility in both pruning questionable tracks and also in identifying   physical phenomena of interest (e.g., endocytosis or quantum dot blinking  \cite{Genovesio2006}) in observed experimental data. 
 Another benefit of our  algorithm is that it  can readily augment  existing algorithms  \cite{Genovesio2006,Serge2008,Jaqaman2008} with negligible additional computational overhead.

%

The two experimental applications presented focus on reliably forming tracks when biomolecules are tagged with 
quantum dot probes.  Quantum dots are attractive because they allow one to track molecules for a long time \cite{Weigel2011,Deutsch2012}, but these probes ``blink'' (hence gap-closing \cite{Jaqaman2008} or other schemes are needed  to follow particles for long times \cite{Serge2008}). 
In the first experimental application presented, we 
quantify how the diffusion coefficient of epidermal growth factor receptors (EGFR), tagged with quantum dots, varies with spatial position in
 the plasma membrane of live COS-7 cells  \cite{Serge2008}.   
Accurately quantifying spatial dependence of the diffusion coefficient and other kinetic parameters  is important in many membrane studies because it provides information about the local molecular environment experienced by tagged particles. For instance, such information can be used to predict the spatial clustering of membrane proteins \cite{Manley2008}. 
 We show how track ambiguity assessment can be used to aid in generating reliable tracks (by pruning questionable associations) for this analysis. 
 To  facilitate algorithmic comparisons, we utilize a publicly available dataset  \cite{Serge2008} in our first experimental application.
  Here, our method is used to process output generated directly from the   ``multiple-target tracing (MTT)" algorithm presented in Ref. 
  \cite{Serge2008}.  

 In order to demonstrate practical utility on larger problems,
  results obtained from modifying the MTT  
  are also analyzed
to demonstrate our algorithm's performance on the same experimental  data in  a scenario where more feasible solutions are considered  (i.e.,   330 putative tracks  are associated to 351 measurements by suitably adjusting  tunable MTT parameters \cite{Serge2008} to relax association gating \cite{BP99}).
Accuracy, timing, and computational complexity results are presented.   We also study an even larger scenario  where many voltage-gated ion channels tagged with quantum dots are tracked.  This larger system demonstrates how the approach introduced can aid in contemporary large scale systems studied in biophysics. In this problem, the tracking algorithm employed, u-track \cite{Jaqaman2008}, requires associating thousands of  track segments in the gap-closing step of u-track
(the problem studied had 39,577 non-zero entries in the  2705$\times$2705 cost matrix defining this problem).  

In addition, we discuss problems arising in other popular  techniques, such as Multiple Hypothesis Tracking (MHT) \cite{Reid1979,BP99,Poore1993,Poore2006} and Interacting Multiple Model (IMM) filters \cite{Blom_IMM_88,Genovesio2006,Jaqaman2008}, when attempting to analyze particles exhibiting complex stochastic motion where accurate 
models are unavailable \emph{a priori}. Large model uncertainty is ubiquitous in SPT applications.
 Particle filters \cite{Godinez2009,Chenouard2009} and MHT trackers \cite{Chenouard2009a} can create biased associations due to assuming inaccurate dynamical models and/or parameters. 
Furthermore, the dynamics of a collection of particles 
can change over time \cite{Fox2011} and hence tracking algorithms appealing to fixed evolution rules can encounter serious problems.  Illustrative examples highlighting the aforementioned problems are presented. 
Techniques for using the algorithm introduced here to improve reliability of SPT algorithms in the presence of model uncertainty are also briefly discussed.

 \section{Background}

 Traditional association problems in target tracking  utilize a variety of discrete optimization routines \cite{BP99,Bertsekas1998}.  Quantifying the ambiguity in such optimization problems often relies on Markov Chain Monte Carlo (MCMC) methods or $k-$best assignment solvers \cite{Kragel2012}.
 However, fast and reliable UQ methods are problematic due to computational concerns outlined in the next section.  For crowded scenarios with poorly understood dynamical models, Probability Hypothesis Density (PHD) filter methods can be leveraged for  tracking the number and position of objects present at any given frame\cite{Mahler2003}, but 
 tracking  and data association 
 (i.e., tracking multiple objects over time) can be problematic with PHD filters if one's goal is to extract kinetic information from putative tracks
 \cite{Panta2004}.  Constructing high quality tracks for kinetic information extraction requires both accurate detection and data association algorithms.

Recently, 
the use of established computational tools from target tracking to solve various association problems
arising  in the analysis of cell biophysics  microscopy data has attracted considerable interest.  
In these applications, reliably forming multiple tracks 
corresponding to numerous physical objects in the field of view of the microscope is of high importance \cite{Kim2006,Manley2008,Danuser2011,Meijering2012}. 
To follow individual molecules over time, researchers in single particle tracking (SPT)  
leverage target tracking \cite{BP99} and computer vision techniques  \cite{Danuser2011}.
For example, one of the early SPT  approaches aiming 
 to extract kinetic information from \emph{in vivo} quantum dot tagged
particles 
can be found in Ref. \cite{Genovesio2006}; the authors employed  a 2D linear assignment solvers and refrained  from using more computationally involved multiple hypothesis tracking (MHT)  techniques due to perceived computational intractability  \cite{Genovesio2006}. 
Established and well-known optimization routines for solving the linear assignment problem, such as the Jonker-Volgenant (JV) algorithm \cite{JValg},  were used in Ref. \cite{Genovesio2006}.  The JV algorithm is also
currently utilized by publicly available MATLAB-based software  to find the 
 optimal global association solution \cite{Jaqaman2008}.  
 Other  publicly available codes  \cite{Serge2008}  overcome computational 
 problems associated with tracking many tagged particles  by employing aggressive  heuristic gating rules that effectively divide the large association problem into several smaller independent local problems (brute force optimization routines are employed after aggressive gating in Ref. \cite{Serge2008}).
More recently, the SPT community has begun developing their own variants  of MHT solvers \cite{Chenouard2009a} 
\footnote{The term SPT is used to denote that individual molecules are tracked over time; many tracking methods with the SPT label do simultaneously track multiple particles}. 
Current comprehensive reviews of existing tools in  SPT can be found elsewhere \cite{Chenouard2009,Danuser2011,Meijering2012}.

Although  
practical MHT solvers (capable of tracking multiple objects in real-time applications) have existed  for some time 
in  target tracking applications \cite{BP99,Poore1993,Poore2006}, the use of these tools in SPT applications 
requires careful thought. For example, in biophysics applications, accurate and reliable models  describing object motion are difficult to determine \emph{a priori}, but many  advanced tracking algorithms require this input.  We provide concrete examples illustrating the problems that can be encountered if one uses tracking approaches depending heavily on \emph{a priori} given models. 
Regardless of the type of algorithm used to form tracks,
 methods for accurately computing the association UQ or ambiguity are helpful is assessing the quality of candidate tracks.  

The approach used here  can leverage existing linear assignment solvers, such as the JV algorithm, or can process the top $k$ scoring association solutions  using modern network flow optimization tools.  
In the results reported, we use  $k$-best linear assignment solvers
allowing one to process rectangular (asymmetric) cost matrices and also consider nodes having infinite capacity \cite{drummond90,Bertsekas1998,Kragel2012}.  
%
%
Our new algorithmic contribution shows how to utilize the information content in the $k$-best solutions and the assumed cost matrix to make better approximations of the uncertainty in candidate association solutions. We also discuss how the tool can assist in other tasks associated with
 large-scale particle tracking problems.  The method we describe is indifferent to the level of sophistication of the assignment solver one utilizes (so researchers without access to advanced discrete optimization codes  can still utilize the tools presented).  For example, we show how the single best global association output of the JV algorithm can be analyzed with the algorithm introduced.


\section{Approach}
\label{sec:approach}
\subsection{Problem Setup}
The goal is to assign $M$ existing tracks, $\{T_m \}^M_{m=1}$,  to $N$ particle measurements, $\{P_n \}^N_{n=1}$, in the current image frame. 
 An arc $s_i$ is defined by a pair of nodes, $(T_{m_i} , P_{n_i} )$; arc $s_i$ can be assigned a likelihood $L_i \equiv L(s_i)$ based on kinetic modeling assumptions \cite{Genovesio2006,Serge2008,Kragel2012}.  Typically, 2D 
 assignment routines work with $\log(L_i)$ \cite{Kragel2012}, we refer to the log likelihoods as the ``scores''.
 Both the set of all feasible arcs and the associated scores are encoded in a cost matrix defining the 2D association problem  \cite{Genovesio2006,Jaqaman2008,Kragel2012}.
 A collection of arcs where each particle is either (i) paired to an existing track  or (ii) assigned to the null-node $\equiv \emptyset$, defines a  hypothesis $\Hypk$; the collection of all feasible hypotheses is denoted by $\Hyps$. Assigning $\emptyset$ to a  particle measurement 
 results in so-called ``track initiation" in the tracking paradigms considered here \cite{Serge2008,Jaqaman2008}.
Note that tracks can also be assigned to $\emptyset$ resulting in a ``missed detection'' event.
As is commonly done in tracking \cite{Kragel2012}, the 
likelihood of any arc associating to $\emptyset$ is set to the value of one (hence making no score contribution).
Assuming associations are statistically independent (a common tracking assumption \cite{Genovesio2006,Serge2008,Jaqaman2008,Kragel2012}),  hypotheses can be ranked 
according to 
$\log\big(L(\Hypk)\big) = \log\big(\prod_{s_i\in \Hypk} L_i\big)$.  
Optimization routines are typically used to extract the top scoring hypothesis in order to update tracks with new particle measurements  \cite{Genovesio2006,Serge2008,Jaqaman2008,Kragel2012}.

Optimization routines do not address ambiguity in assignments. 
Vastly different hypotheses can have effectively identical scores. 
 In tracking, standard approaches attempt to compute the exact hypothesis probability, $P(\Hypk) = \frac{L(\Hypk)}{\sum_{\Hypk \in \Hyps}{L(\Hypk)}}$,
 to quantify this problem  \cite{Kragel2012}; for example, $P(\Hypk)> 0.95$ may suggest an ``unambiguous association" solution in the problem under study.
Accurately quantifying that specific arc associations are likely correct (as opposed to estimating $P(\Hypk)$) is often more informative in biological SPT applications, since multiple hypotheses may contain an arc of scientific interest.  
Denote an arc for which we want to compute the association probability by $s$ (we refer to this as the ``target arc"); assume that this target arc contains two non-null nodes. Define $\HypsFix{s}$ to be the set of all hypotheses containing $s$ and let $\HypsHid{s}$ be the set of all hypotheses that do not contain $s$. The arc ambiguity can then be written as: 
$P(s) = \frac{\sum_{\Hypk \in \HypsFix{s}}{L(\Hypk)}}{\sum_{\Hypk \in \Hyps}{L(\Hypk)}}$. 

For moderate problem sizes, simple combinatorics makes brute force approaches to computing exact arc or hypothesis probabilities computationally intractable or problematic.  A traditional approach used to circumvent problems introduced by an exponential number of members in $\Hyps$ is to define a proper subset of hypotheses, 
$\HypsSample \in \Hyps$, and then compute $\hat{P}(\Hypk) = \frac{L(\Hypk)}{\sum_{\Hypk \in \HypsSample}{L(\Hypk)}}$
 \cite{Kragel2012}.  
 The hat denotes an approximated probability.  The analogous arc probability approximation is
 $\hat{P}(s) = \frac{\sum_{\Hypk \in \HypsSample \cap \HypsFix{s}}{L(\Hypk)}}{\sum_{\Hypk \in \HypsSample \cap \Hyps}{L(\Hypk)}}$.  We label the above truncation schemes for computing $\hat{P}(\Hypk)$ or $\hat{P}(s)$ as the ``Traditional Method'' \cite{Kragel2012}.  
Our method aims to improve the accuracy of $\hat{P}(s)$ by making better use of the information contained in
 the cost matrix and $\HypsSample$  in situations where complete enumeration of $\Hyps$ is not possible or desirable.

\subsection{The Correspondence Method}

The input to the our algorithm consists of the cost matrix defining the 2D association problem, the target arcs  of interest $ \{ s_t\}_{t=1}^T$, and a set $\HypsSample$ of candidate
association solutions.  Solutions can be generated by various means.  This article utilizes  $k$-best solvers \cite{Danchick2006,Kragel2012}, i.e., $\HypsSample \equiv \{H_1,H_2,\ldots H_k \}$ such that $L(H_1)\ge L(H_2) \ge \ldots L(H_k)$. The output is $ \{ \hat{P}(s_t)\}_{t=1}^T$, where 
$\hat{P}(s_t)$  is an estimate of the exact arc probability, $P(s_t)$.
 Ambiguity estimates  \emph{can be computed using only one candidate solution, so the method can be utilized by existing trackers where the association solver cannot be readily modified or replaced}.  After presenting the general formalism in this subsection, we illustrate the method on a concrete small toy simulation example to clarify the 
 basic intuition behind the algorithm
 in the next subsection (the formalism presented here facilitates the method's computational implementation outlined in the Appendix). 

For a given target arc, $s$, our method partitions 
the hypothesis space into clusters, such that each cluster $C^{s}_{\Hypk}$ corresponds to  a unique $\Hypk \in \HypsFix{s}$ (this key feature inspired the algorithm's name, i.e., the \emph{Correspondence Method}). Recall that an arc $s$ connects generic nodes $a$ and $b$; below it is convenient to explicitly represent nodes in an arc via $ab\equiv s$. For each $\Hypk \in \HypsFix{s}$, the cluster $C^{ab}_{\Hypk}$  is defined by:
$\{ \Hypk-ab-xy+xb+ay : 
xy \in \Hypk \ s.t. \ (xb \in A \wedge ay \in A)
\}$
where $A$ is the set of all feasible arcs. That is, cluster $C^{ab}_{\Hypk}$ is generated by swapping $ab$ with each of the other arcs in $\Hypk$, one at a time.  

For each hypothesis $\Hypk \in \HypsSample$, we determine which cluster $\Hypk$ belongs to. If $\Hypk \in \HypsFix{s}$, then $\Hypk$ trivially belongs to cluster $C^{s}_{\Hypk}$. If $\Hypk \in \HypsHid{s}$, we let $a^{\prime} = adj_{\Hypk}(a)$, where 
node $adj_{\Hypk}(a)$ denotes the node  associated to  $a$ in hypothesis $\Hypk$ 
(i.e., the other node of the arc  in $\Hypk$ containing $a$, or $\emptyset$ if no such arc exists), and let $b^{\prime} = adj_{\Hypk}(b)$. Then, we have $\Hypk \in C^{ab}_{(\Hypk-aa^{\prime}-b^{\prime}b+b^{\prime}a^{\prime}+ab)}$. However, if $b^{\prime}a^{\prime} \notin A$, then $\Hypk-aa^{\prime}-b^{\prime}b+b^{\prime}a^{\prime}+ab$ does not exist. In this case, the hypothesis is in a singleton cluster. Cluster formation is illustrated graphically in the Appendix.
Each cluster's likelihood is computed and subsequently re-weighted to generate $\hat{P}(s)$.  The re-weighting procedure 
is outlined next.

For each cluster $C$, define the  \emph{cluster probability}:
\begin{align}
\label{eq:clustp}
Q_{s}(C)=\frac{\sum_{\Hypk  \in (C \cap \HypsFix{s})}{L(\Hypk)}}{\sum_{\Hypk  \in C}{L(\Hypk)}}
\end{align}
 Note that $Q_{s}(C)$ is the fraction of the probability mass in the cluster that consists of the hypotheses in $\HypsFix{s}$. 
Next, a weighted cluster sample $\CSample=\{C(\Hypk) | \Hypk \in \HypsSample\}$ is constructed where $C(\Hypk)$ is the cluster containing $\Hypk$.  Each cluster $C \in \CSample$ has weight $w(C)=\sum_{\Hypk \in C \cap \HypsSample}L(\Hypk)$. 
Our estimate of the association probability of $s$ given such a sample $\ClusterSample{s}$ is:
\begin{align}
\label{eq:phat}
\hat{P}(s) = \frac{\sum_{C \in \ClusterSample{s}}{Q_{s}(C)w(C)}}{\sum_{C \in \ClusterSample{s}}{w(C)}}
\end{align}

\subsubsection{Illustrative Example on a Toy Problem}
\begin{figure}[htb]
\center
\centering
\begin{minipage}[b]{.4\linewidth}
\def\pw{.99}
  \includegraphics[width=\pw\textwidth]{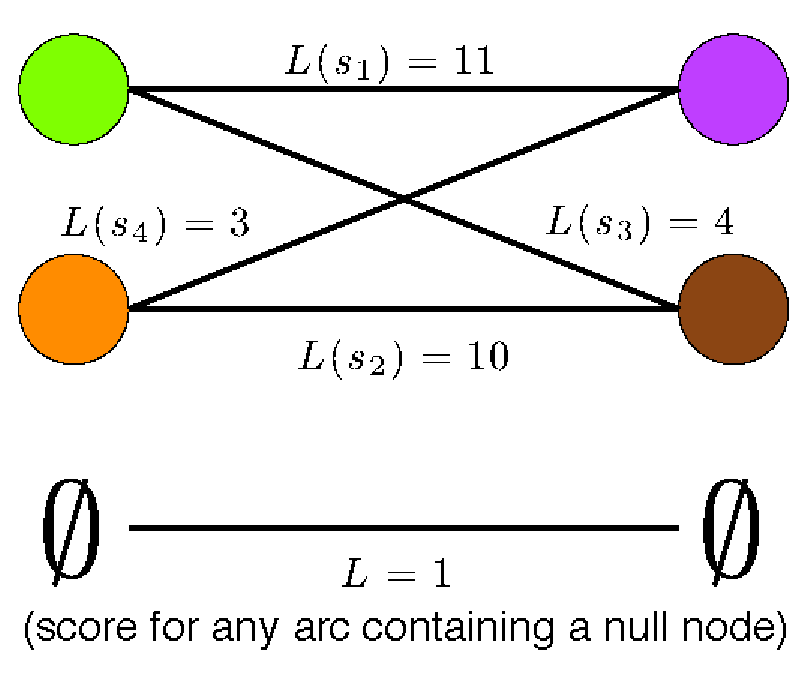}
  \end{minipage}
\centerline{\footnotesize }
\caption{
\footnotesize 
Toy problem setup.  Likelihoods of feasible arcs displayed. Recall that any arc containing the null node has a likelihood score of unity, and thus a cost of zero. 
 }
\label{fig:illustrativeEg}
\end{figure}

To concretely demonstrate  the approach, we illustrate a toy $2\times2$ association 
problem where exact  computation is possible (only 7 feasible hypotheses exist).  Figure \ref{fig:illustrativeEg} displays the arcs and the associated likelihood score for the problem.

First suppose, that one has access to an optimization routine  providing only the single best global hypothesis (i.e., $k=1$).  In this problem, $H_1=\{s_1,s_2\}$ and  $L(H_1)=110 =11\times10$.   Recall that the cluster scheme is defined by a given arc $s_i$ and $\HypsSample$.  
So to compute $\hat{P}(s_1)$, one simply forms the cluster members defined by arc $s_1$.
Figure \ref{fig:res2a} displays all of the cluster corresponding to analyzing  $s_1$; in this toy problem, only two clusters exist. Since $s_1 \in H_1$, one of the clusters is labeled as $C^{s_1}_{H_1}$.     
When
$\HypsSample$ only contains one member, the $\hat{P}(s_1)$ computation is straightforward;  one simply computes $
Q_{s_1}(C^{s_1}_{H_1})=\frac{110}{110+12+10}\approx 0.8333$ (the denominator comes from scoring the three hypotheses  making up cluster $C^{s_1}_{H_1}$; the members of this cluster are shown in  Fig. \ref{fig:res2a}).
Given that only one hypothesis was considered, it can be observed from Eqn. \ref{eq:phat} that $\hat{P}(s_1)=Q_{s_1}(C^{s_1}_{H_1})$.

\begin{table} [htb]
\center
\caption{\label{tab:illeg}  Arc UQ as a function of $k$ for Traditional (Trad.) and the Correspondence Method (CM).\baselineskip=10pt  }
 \center
\begin{tabular}{|l|*{8}{c|}} \hline 

  & \multicolumn{2}{c|}{ $s_1$ } & \multicolumn{2}{c|}{ $s_2$ } & \multicolumn{2}{c|}{ $s_3$ } & \multicolumn{2}{c|}{ $s_4$ } \\ \hline

\# $k$-best  & Trad. & \color{red} CM &  Trad. & \color{red} CM &  Trad. & \color{red} CM &  Trad. & \color{red} CM \\ \hline
k=1 & 1.0000  & {\color{red}0.8333 } & 1.0000  & {\color{red}0.8271 } & 0.0000  & {\color{red}0.0960 } & 0.0000  & {\color{red}0.0952 } \\ \hline

k=2 & 0.9016  & {\color{red}0.8333 } & 0.9016  & {\color{red}0.8271 } & 0.0984  & {\color{red}0.0960 } & 0.0984  & {\color{red}0.0952 } \\ \hline

k=3 & 0.9098  & {\color{red}0.8123 } & 0.8271  & {\color{red}0.8271 } & 0.0902  & {\color{red}0.1008 } & 0.0902  & {\color{red}0.0973 } \\ \hline

k=4 & 0.8462  & {\color{red}0.8138 } & 0.8392  & {\color{red}0.8081 } & 0.0839  & {\color{red}0.1045 } & 0.0839  & {\color{red}0.0989 } \\ \hline

k=5 & 0.8231  & {\color{red}0.8074 } & 0.8163  & {\color{red}0.8012 } & 0.1088  & {\color{red}0.1058 } & 0.0816  & {\color{red}0.0988 } \\ \hline

k=6 & 0.8067  & {\color{red}0.8028 } & 0.8000  & {\color{red}0.7963 } & 0.1067  & {\color{red}0.1056 } & 0.1000  & {\color{red}0.0992 } \\ \hline

k=7 & 0.8013  & {\color{red}0.8013 } & 0.7947  & {\color{red}0.7947 } & 0.1060  & {\color{red}0.1060 } & 0.0993  & {\color{red}0.0993 } \\ \hline

\end{tabular}
\end{table}

The case where $k>1$ hypothesis  requires one to utilize the cluster weighting scheme.  Suppose we have $k=3$;  in the toy problem under consideration, one augments $H_1$ with $H_2=\{s_3,s_4\}$ ($L(H_2)=12=4\times 3)$ and $H_3=\{s_1\}$ ($L(H_3)=11$). $H_2$ is actually a member of cluster $C^{s_1}_{H_1}$ (this cluster probability was already computed for $k=1$), hence the weight of this cluster coupled with the given  $k=3$ top hypotheses increases  to $w(C^{s_1}_{H_1})=122=110+12$.  
 $H_3$ is a hypothesis containing the target arc $s_1$ and subsequently forms a new cluster;
 from this new cluster, $C^{s_1}_{H_3}$ (shown in Fig. \ref{fig:res2a}), one can compute the corresponding cluster probability $Q_{s_1}(C^{s_1}_{H_3})=\frac{11}{19}$; the numerator comes from $L(H_3)$ and the denominator comes from computing the sum of scores  in all of the hypotheses in $C^{s_1}_{H_3}$ ($19 = 11 + 4+ 3 +1$).
   Since $H_3$ is the only member of cluster $C^{s_1}_{H_3}$ present in the top $k=3$ solutions, the weight of this cluster is also $L(H_3)=11$.  So our method's estimate for $k=3$, utilizing Eqn. \ref{eq:phat}, is $\frac{11 \times \frac{11}{19}+122\times\frac{110}{132}}{122 + 11}\approx 0.812294$.  Table \ref{tab:illeg} displays results for all other arcs and $k$ values.

\subsection{Consistency of the Correspondence Method}
Our method introduces zero bias when $\HypsSample=\Hyps$. That is, we show that $\hat{P}(s)$ is precisely equal to the true $P(s)$ when $\HypsSample=\Hyps$. In this case, one has:
\begin{equation}
 Q_{s}(C)w(C) = \sum_{\Hypk \in (C \cap \HypsFix{s})}{L(\Hypk )}, 
\end{equation}
 and 

\begin{align}
 \sum_{C \in \ClusterSample{s}}{Q_{s}(C)w(C)} = & \sum_{\Hypk  \in \HypsFix{s}}{L(\Hypk )} \\
 \newline \sum_{C \in \ClusterSample{s}}{w(C)} = & \sum_{\Hypk  \in \Hyps}{L(\Hypk )},  
\end{align}
so plugging the above into Eqn. \ref{eq:phat} shows that $\hat{P}(s) = P(s)$.

\subsection{Correspondence Method Summary and Comments}
The main ideas underlying the Correspondence Method have now been presented. Additional  technical details utilized in practical computation and a complexity analysis 
are contained in the Appendix.  The essence of the idea is to  create additional feasible hypotheses (i.e., associations solutions)
from given candidate solutions and then utilize the original and supplemental information to improve
arc ambiguity UQ computations by systematically combining the information to produce arc ambiguity estimates.
The approach is guaranteed to be unbiased after enumeration of all hypotheses (as shown above). 
The traditional $k-$best method shares this property  \cite{Kragel2012}.
However, we provide several large-scale examples illustrating why the convergence of the Correspondence Method is much 
faster that the traditional $k-$best method \cite{Kragel2012}.
The Correspondence Method  shows great promise in scenarios where one can only compute a small fraction of the top scoring feasible hypotheses due either to time or memory constraints.  In situations where many of the top scoring hypotheses overlap in multiple arcs, our method provides a means for extracting more reliable arc UQ estimates implied by both (i) the  top 
$k$ solutions directly outputted by the discrete $k$-best solver and (ii) associations implied by hypotheses  constructed in the cluster
formation phase of the Correspondence Method. The rapid enhanced sampling of hypothesis space  afforded by the cluster probability computation is the reason for the method's improved accuracy.   

\section{Results and Discussion}
\label{sec:res}

\begin{figure*}[htb]

\begin{minipage}[b]{.225\linewidth}
  \centering
  \def\pw{1}
\def\pww{.3}
  \centerline{\footnotesize    Tracks   \hfill Particles}
  \includegraphics[width=\pw\textwidth]{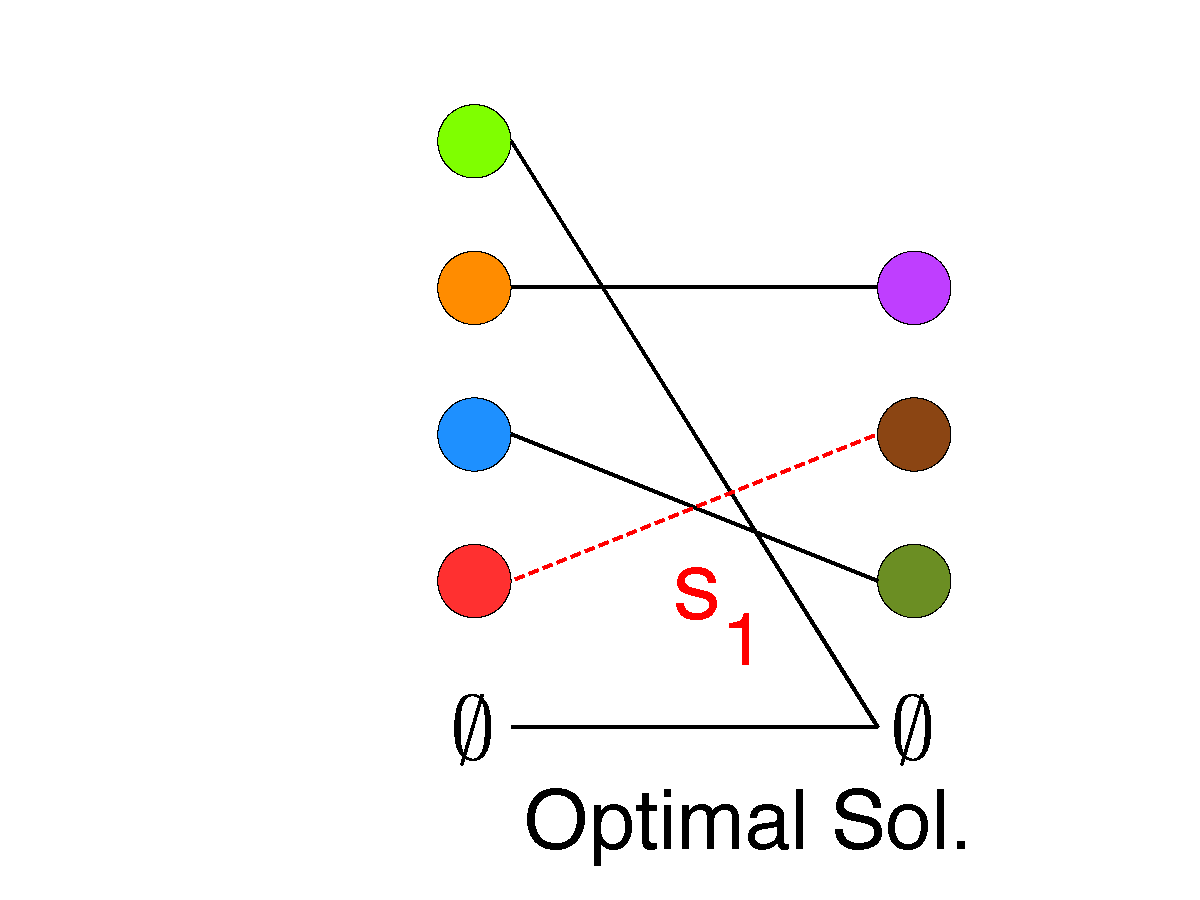} \\
   \centerline{\footnotesize   Tracks   \hfill Particles}
    \includegraphics[width=\pw\textwidth]{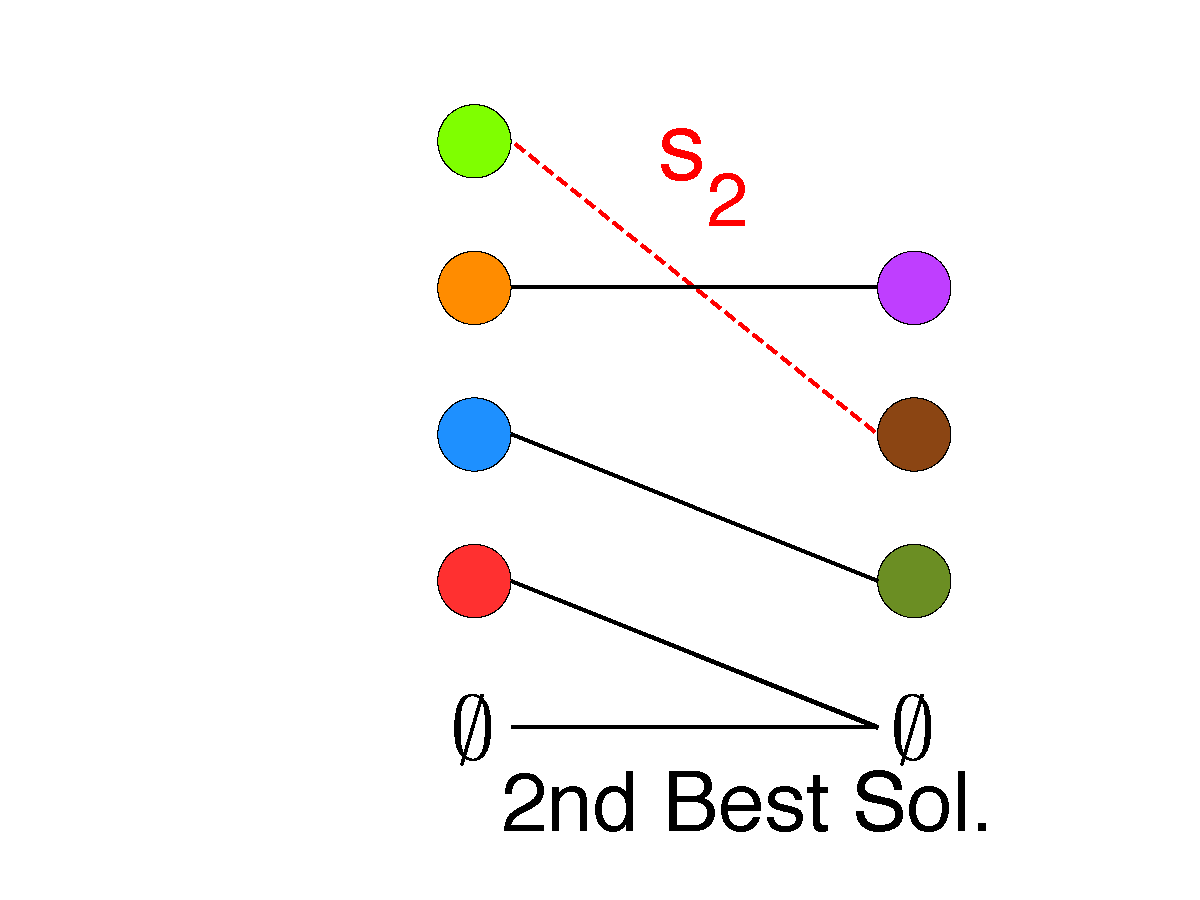}
  
\end{minipage}
\hfill
\begin{minipage}[b]{.775\linewidth}
  \centering
  \def\pw{.425}
\def\pww{.25}
  \includegraphics[width=.44\textwidth]{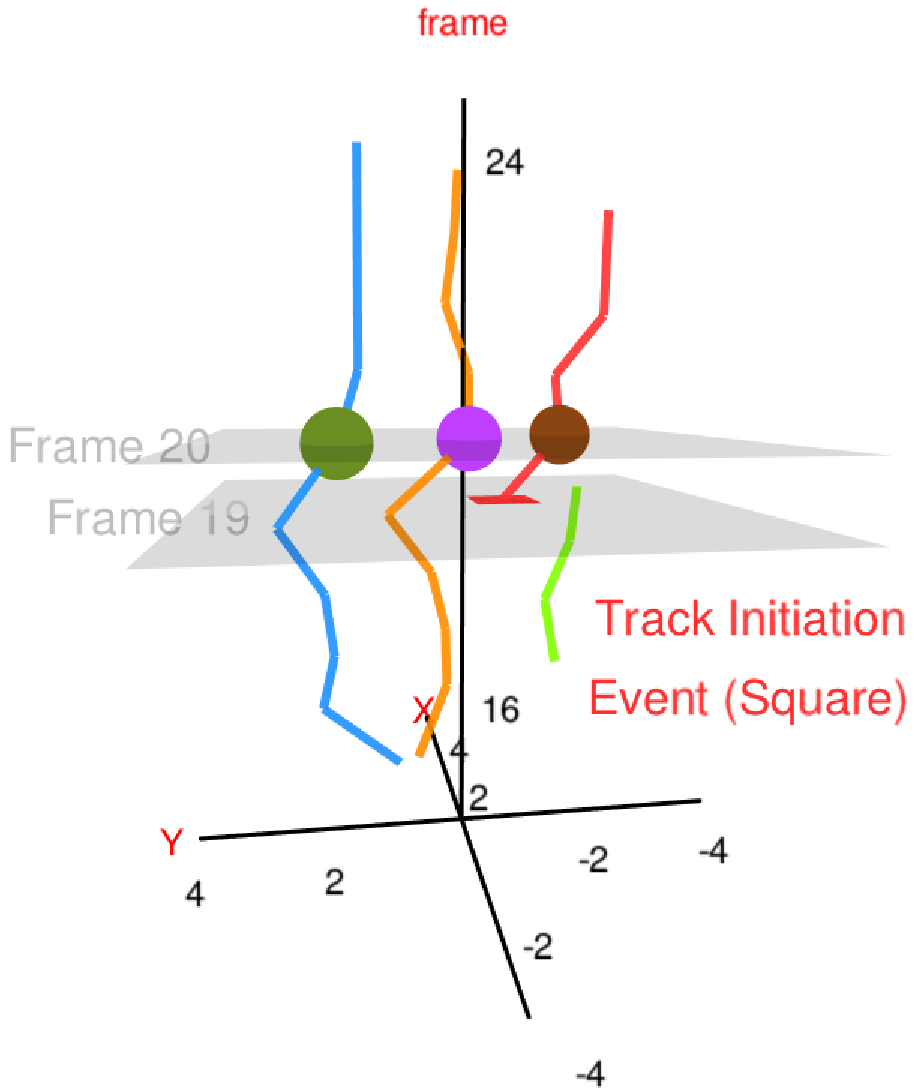}
  \hfill
  \includegraphics[width=.4475\textwidth]{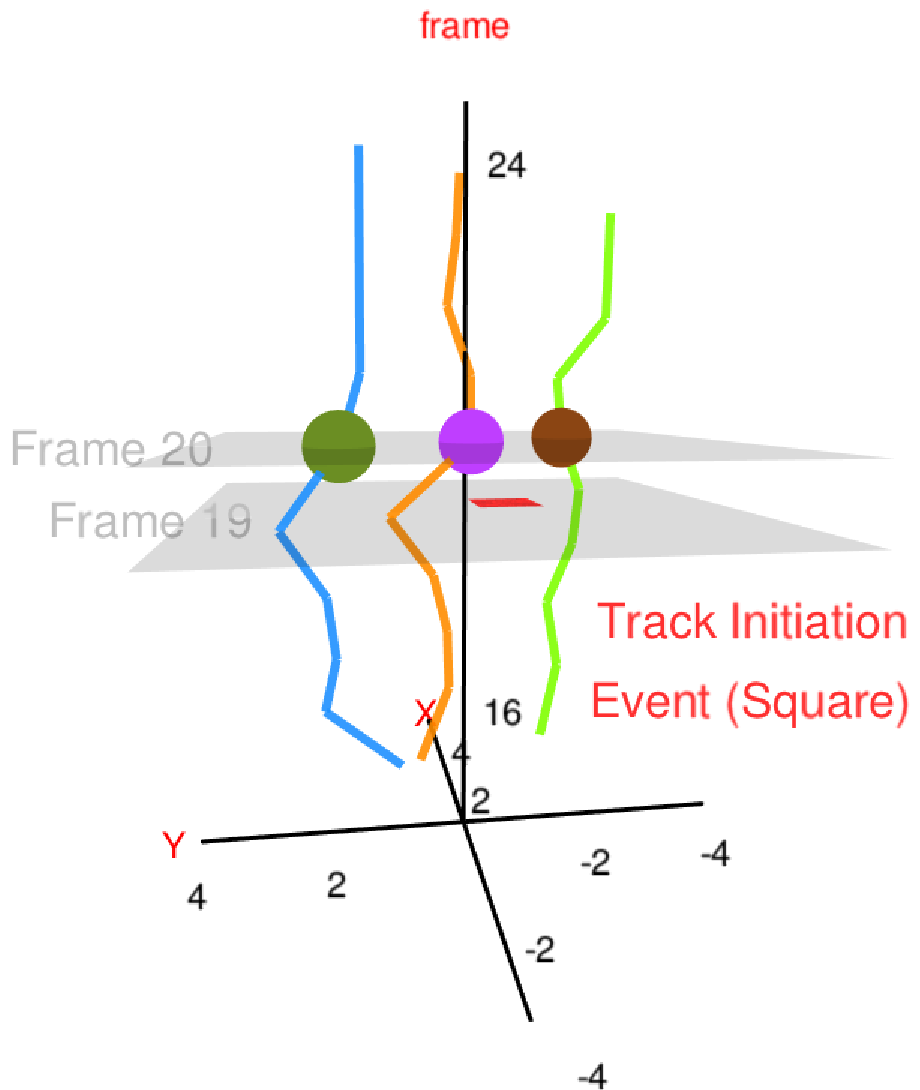}  
  \hfill
\end{minipage}

\caption{
\footnotesize Two hypotheses connecting frames 19 and 20 (left panel). 
Results corresponding to selecting each hypothesis  (middle panel [tracks resulting from selecting Optimal Soln.] and right panel [2nd Best]); tracks denoted by lines and measurements  by spheres.  The two hypotheses differ only by arcs $\color{red} s_1$ and $\color{red}  s_2$. 
Note:  Only 3 candidate particle measurements are available to pair with 4  tracks;  also only a portion of the tracks are plotted (data for frames 1-15 not shown).}
\label{fig:res1}
\end{figure*}

Figure \ref{fig:res1}  
displays the two top scoring hypotheses in frame 20 from Ref. \cite{Serge2008}'s data. Four tracks formed from frames 1-19 need to be assigned to three measurements. The top two association solutions differ only by swapping which track is assigned  to the null node (i.e., a putative missed detection). The tracks in this figure 
were formed using the tracker presented by Serg\'{e} \emph{et al.} \cite{Serge2008};  this tracker
only considers tracks and measurements in a small local spatial area (default parameters were used except the parameter ``\url{Boule_free}" \cite{Serge2008} was increased from 3.5 to 4.0  to reduce the severity of heuristic gating). More specific track and measurement information characterizing the algorithm is provided in the Appendix.

\begin{table}[ht]
  \newcolumntype{C}{>{\centering\arraybackslash}m{1.2695cm}<{}}
  \newcolumntype{T}{>{\centering\arraybackslash}m{1.4cm}<{}}
   \newcolumntype{L}{>{\arraybackslash}m{1.6cm}<{}}
   \center
  \caption{\footnotesize{``Traditional'' denotes the arc ambiguity estimate discussed in Sec. \ref{sec:approach} and ``CM'' indicates Correspondence Method.  }}
  \label{tab:PA1}
\begin{tabular}{|L|T|C|C||T|C|C|}
\hline
\multirow{2}{*}{ \textbf{\color{methcolor} Method}}  & \multicolumn{3}{|c|}{ \textbf{ \color{blue} Local}  $ \color{blue}  (4\times 3)$ }& \multicolumn{3}{|c|}{  \textbf{ \color{blue}Global}  $\color{blue}  (330\times 351)$} \\
& {\footnotesize comp. time [s]} & $\hat{P}({\color{red} s_1})$ & $\hat{P}({\color{red} s_2})$ & {\footnotesize comp. time [s]} & $\hat{P}({\color{red} s_1})$ & $\hat{P}({\color{red} s_2})$ \\

\textbf{\color{kcolor} k=1} & & & & & &\\
Traditional & 1$\times10^{-3}$ & 1 & 0 & 4$\times10^{-1}$ & 1& 0 \\
CM           &  2$\times10^{-3}$ & 0.626 & 0.374 & 4$\times10^{-1}$ & 0.626 & 0.374 \\
\textbf{\color{kcolor} k=4} & & & & & &\\
Traditional & 2$\times10^{-3}$ & 0.621 &  0.379 & 4$\times10^{-1}$ & 0.769   & 0.231  \\
CM  & 3$\times10^{-3}$  & 0.617 & 0.383 & 5$\times10^{-1}$ & 0.626 & 0.374 \\
\textbf{\color{kcolor} k=50}& & & & & &\\
Traditional & 8$\times10^{-3}$ & 0.611 & 0.388  & 5$\times10^{-1}$ & 0.646  &  0.355  \\
CM  & 9$\times10^{-3}$  & 0.611 & 0.388 & 6$\times10^{-1}$ & 0.623 &  0.376 \\
\hline
\textbf{\color{kcolor} \textbf{Truth}} &-&$\color{red} \textbf{0.611}$&$\color{red} \textbf{0.388}$&-& \color{red}\textbf{0.615$^\dagger$} &  \color{red}\textbf{0.385$^\dagger$} \\
\hline
\end{tabular}
\\  \ \hfill \footnotesize {$^\dagger$ Truth computed using ``Traditional" $k-$best estimate  \cite{Kragel2012} with $k=10^4$.} 
\label{tab:timing}
\end{table}



 In Tab. \ref{tab:timing}, we approximate the uncertainty associated with the  arcs distinguishing the top two hypotheses using various methods.  In the ``local $k=1$" runs, we utilize the  direct output of Serg\'{e} \emph{et al.}'s  tracker \cite{Serge2008}. An advantage of analyzing the local association problem (i.e., the formulation using aggressive gating
 resulting in several disjoint local  association neighborhoods \cite{Serge2008})  is
 that the local  $4\times3$ problem contains only 73 $\big(=\sum_{a=0}^3{4\choose a} \frac{3!}{(3-a)!}\big)$ 
 hypotheses, so exact arc ambiguity can  be readily computed (see bottom row in Tab.  \ref{tab:timing}). The arcs in the best local solution provide an accurate estimate of these probabilities whereas the traditional methods do not even consider the possibility of incorrect associations (see ``$k=1$ Traditional'').
We also used a $k-$best solver to construct additional hypotheses to serve as input to both  the Traditional  Method     and the Correspondence Method;  the enhanced accuracy of using $k>1$ is quantified.

 \begin{figure}[htb]
\center
\centering
\def\pw{.65}
  \includegraphics[width=\pw\textwidth]{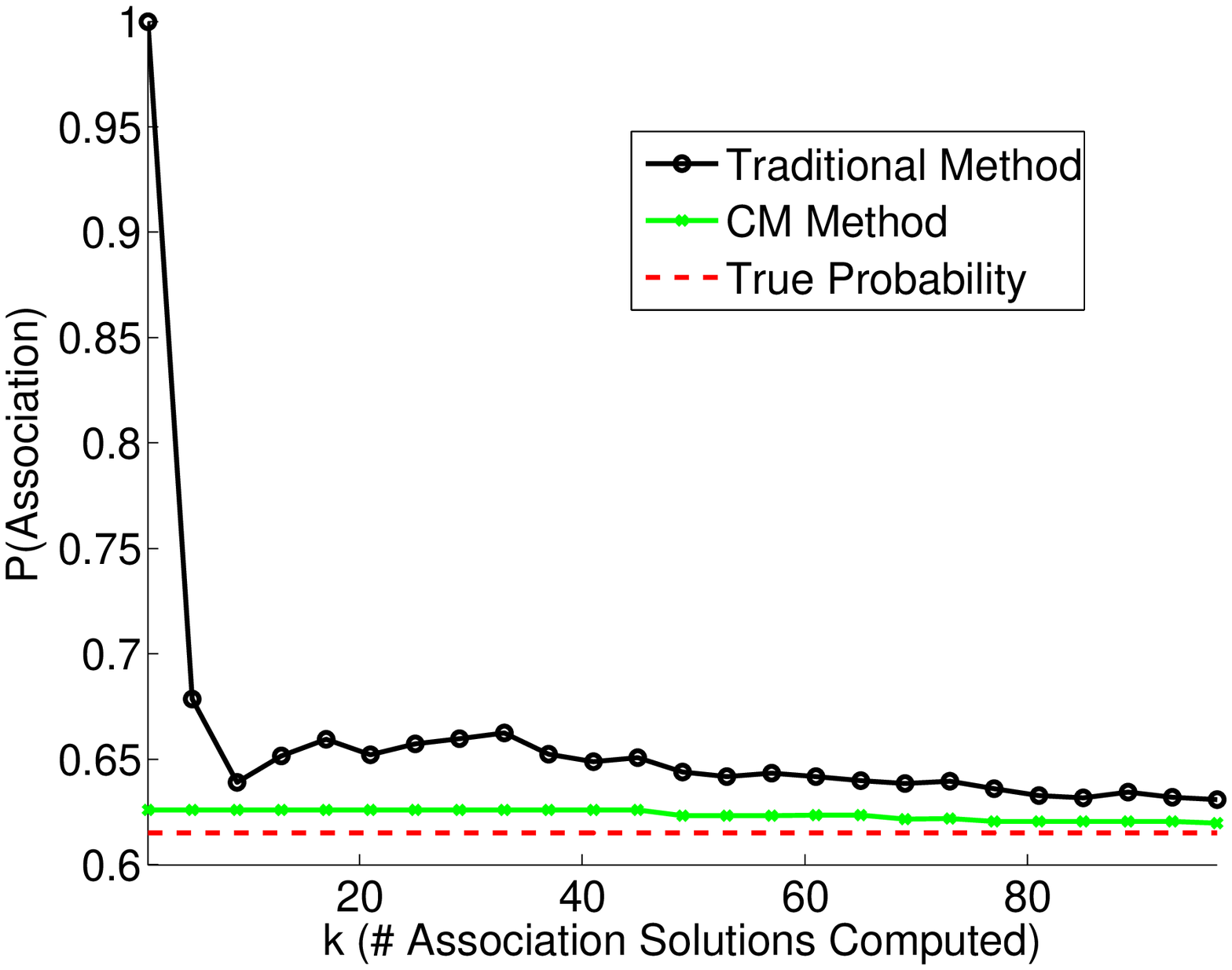}
\centerline{\footnotesize Rate of convergence of $\hat{P}(s_1)$ for 
 $330\times 351$ association problem shown in Table \ref{tab:PA1}.}
\caption{
\footnotesize 
 }
\label{fig:rate}
\end{figure}

\subsection{Global Performance on Larger 2D Association Problems}

 Table  \ref{tab:timing} and Fig. \ref{fig:rate} 
 display results where, in the same frame, we completely relax the aggressive gating 
 of Serg\'{e} \emph{et al.}'s  tracker and construct a dense global association problem (i.e., to extend frame 19, 330 tracks need to be associated with 351 measurements). 
 Note that in the table, the ``truth" cannot be computed precisely in this larger problem ($k=10^4$ results serve as the true arc association probability). Figure  \ref{fig:rate} displays the rate
 of convergence for  $k\in [1,100]$.
 %
 The computational timing results in Tab.  \ref{tab:timing} (measured on a single core using Matlab and a C++ ``mex file"-based implementation of our algorithm and the $k-$best assignment solver) demonstrate that the global association problem \emph{with track ambiguity computations} can be readily handled. 
 If one can easily switch out association solvers, it is clearly worth the effort, but an appealing feature of the Correspondence Method is that it can compute accurate association uncertainty using standard SPT output without modifying the assignment solver. 


The top association implies that a recently initiated 
track (initiation point shown via red square) is assigned a new measurement whereas a longer living track is 
unassigned. The second best hypothesis implies that the newly formed track has a missed detection in the frame immediately after track initiation implying that the putatively confirmed measurement  was likely just fluorescence noise (i.e., a false alarm).  Deciding between the top two hypothesis  is effectively a ``coin flip" given the ambiguity scores. 
Rationally deciding which hypothesis to believe (if any)  requires subject matter expertise and/or tools analyzing multiple frames of data, such as a MHT tracker. This is an example of a case where MHT technology can help;  results presented later show that MHT trackers do not always resolve the ambiguity issue.  In any case, the Correspondence Method  allows one to automatically flag  cases where large ambiguity resides with minimal overhead in large scale problems.

\begin{figure}[htb]

\begin{minipage}[b]{1.0\linewidth}
  \centering
    \centerline{\footnotesize $2D$-Diffusion Coefficient Magnitude $[\mu m^2/s]$}
  \centerline{\includegraphics[width=4.5cm]{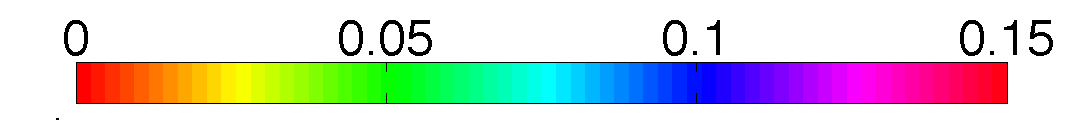}}

\end{minipage}

\begin{minipage}[b]{.42\linewidth}
  \centering
    \def\pw{.99}
    \centerline{\includegraphics[width=\pw\textwidth]{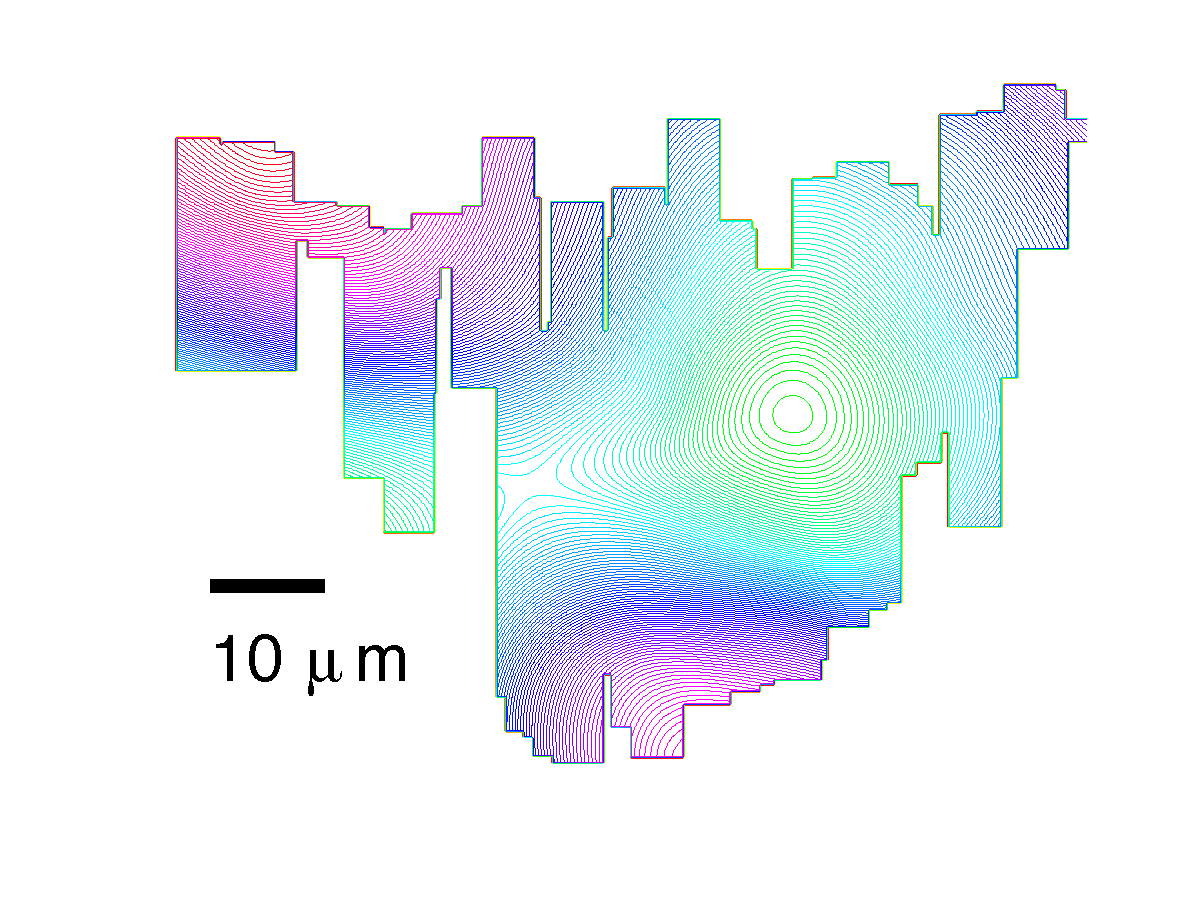}}
  \centerline{\footnotesize  (a) Accepting all tracks}\medskip
\end{minipage}
\hfill
\begin{minipage}[b]{0.42\linewidth}
  \centering
   \def\pw{.99}
    \centerline{\includegraphics[width=\pw\textwidth]{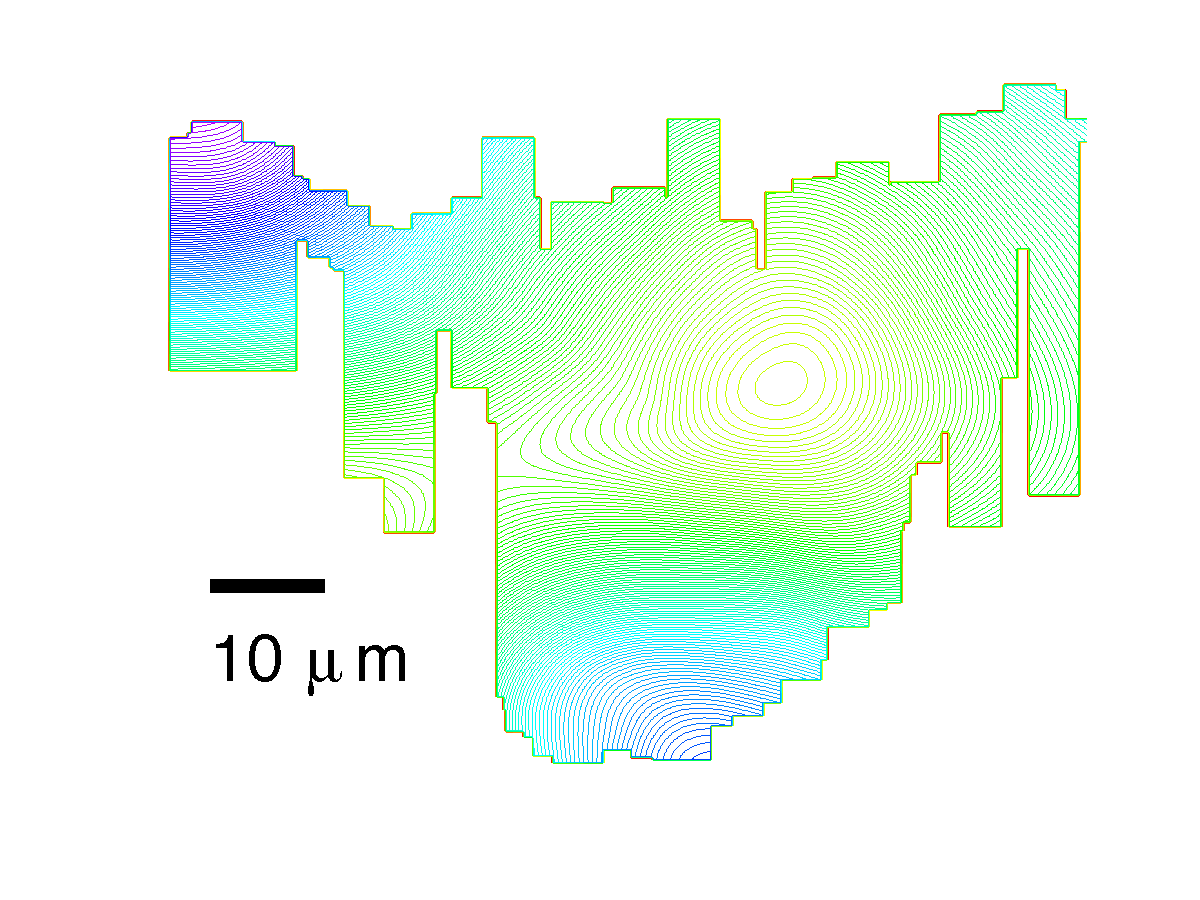}}
  \centerline{\footnotesize  (b) Pruning uncertain tracks}\medskip
\end{minipage}

\caption{
\footnotesize Estimated spatial dependence of diffusion coefficient.}
\label{fig:res2}
\end{figure}

Next, we reanalyzed all the experimental tracks from Ref. \cite{Serge2008}. 
  The local tracker described above (i.e., the tracking algorithm  parameter settings resulting in the aggressive ``local'' gating) was used to generate tracks.   This output was  
 then used to estimate a collection of
 diffusion coefficients by fitting a linear stochastic differential equation 
 via maximum likelihood estimation (MLE) to each individual track;  Fig. \ref{fig:res2} plots the estimated 2D diffusion coefficient as a function of space (MLE and technical surface fitting details are deferred to the Appendix).  In the left panel,  345 tracks had their diffusion coefficient estimated 
(only tracks having greater than 25 time ordered entries were fit by MLE).
Any  track  involved in an assignment whose top arc had a net probability less than 95\% was flagged as suspicious and the tracks in the hypothesis were pruned from the collection; this resulted in removing 121 tracks \footnote{The default settings of the MTT tracker resulted in 1-4 arcs per local association hypothesis (recall this algorithm is intended to divide the problem into many small association problems by aggressive gating \cite{Serge2008,BP99}). The difference between results obtained by omitting all tracks vs. only those in questionable arcs was found to be insignificant.  In the u-track example, the advantages of taking the arc view become readily apparent. }.   
%
 The population statistics 
 of the full vs. pruned diffusion coefficient estimates differed substantially, e.g., median  0.0457 vs. 0.0264 $\mu m^2/s$. 
 This suggests that uncertain tracks biased the estimate (e.g., particles received artificially large fluctuations due to incorrect associations). Alternatively, this could imply that certain regions of high diffusivity are too crowded for reliable track formation. 
 The latter explanation is weakened since the similar spatial regions of higher  diffusivity can still be observed; the magnitude is simply lower when uncertain tracks are pruned.

In this previous example, we aggressively / conservatively pruned tracks.  Any track 
containing an arc possessing  $\hat{P}(s)$ was completely ignored in the downstream diffusion coefficient analysis. 
One can consider alternative pruning strategies;
since u-track    
utilizes both frame-to-frame  as well as a ``gap-closing'' (GC) associations  \cite{Jaqaman2008} (the latter type of association is discussed in the next subsection), these features facilitate removing ambiguous measurements from a collection of tracks without necessarily sacrificing track length.
 The Correspondence Method can assist in more carefully pruning ambiguous 
 segments of tracks (hence retaining more data) in both phases of u-track if minor modifications are made to the algorithm.  
 However, before presenting these results, we provide an example illustrating 
 some problems   biological applications face when tracking multiple particles when
 stochastic models are unknown \emph{a priori}. We also use the example to 
 sketch the ideas behind both phases of u-track \cite{Jaqaman2008}.

\subsection{Dangers of Model Uncertainty in SPT Applications}
\begin{figure}[htb] 
\center
\centering
\def\pw{.475}
  \includegraphics[width=\pw\textwidth]{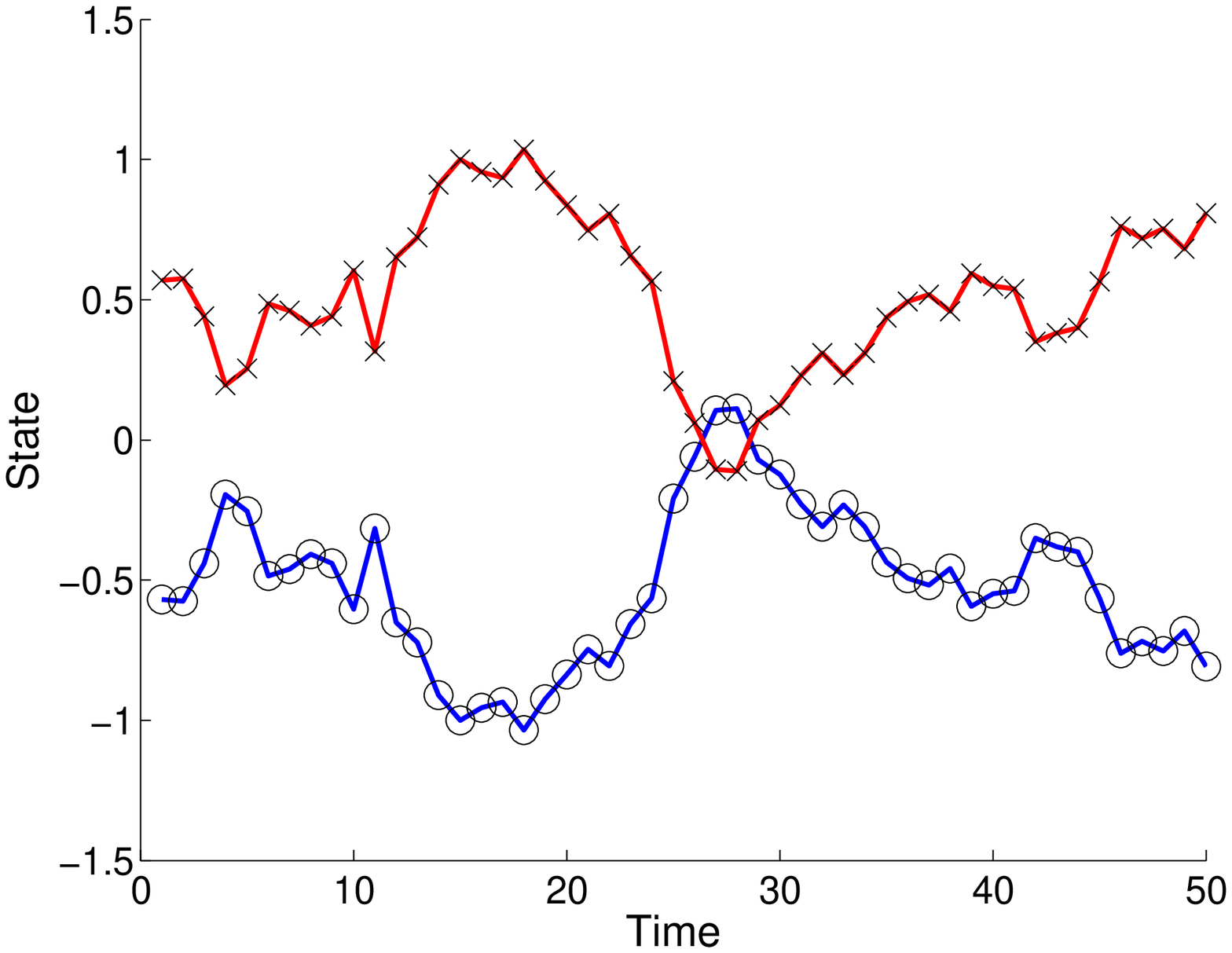}
  \includegraphics[width=\pw\textwidth]{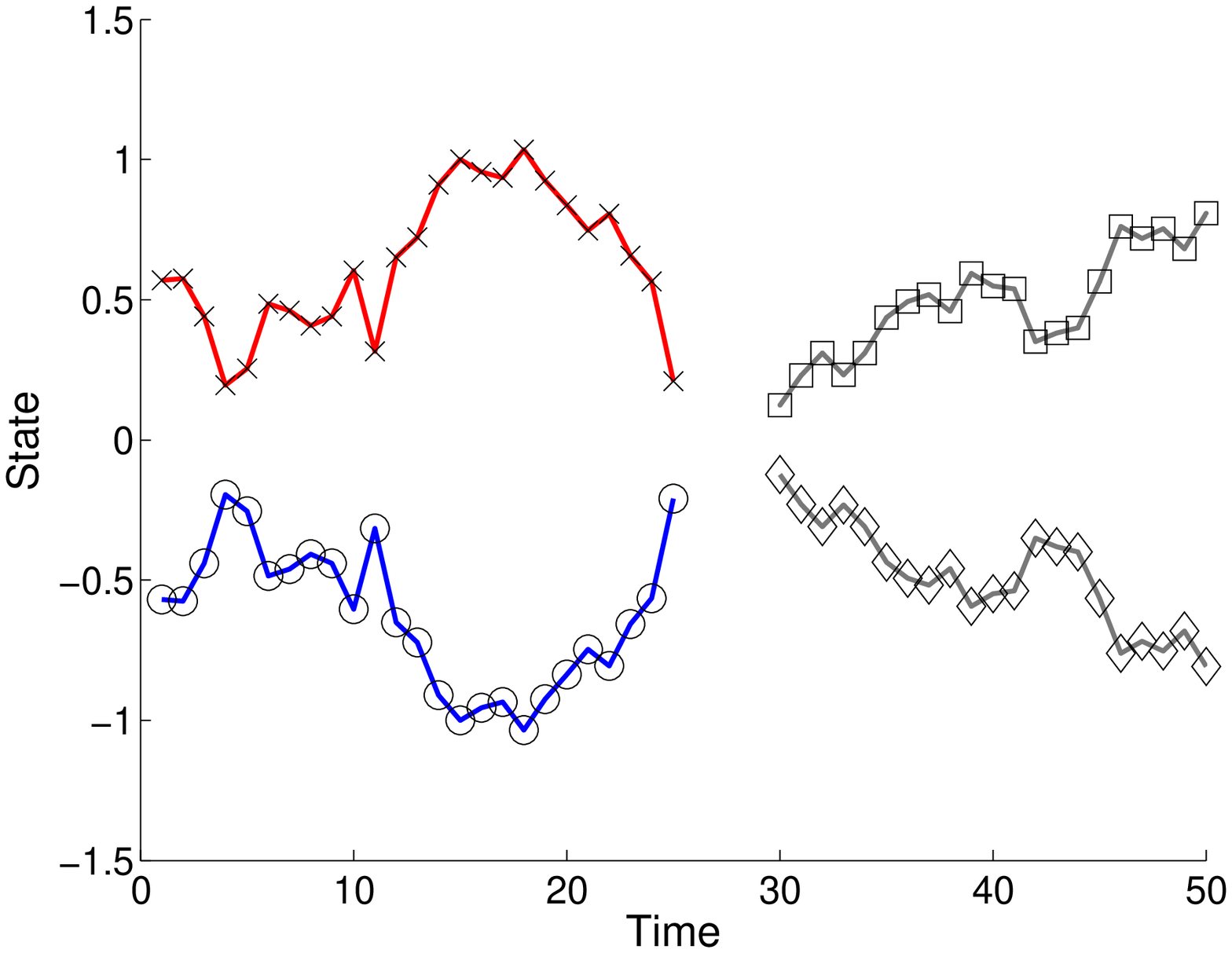}
\centerline{\footnotesize }
\caption{Illustration of track crossing ambiguity.  The left panel displays the true trajectories.  However, for this data generating process, there is no statistical evidence
available for making unambiguous associations given only a collection of discretely sampled position coordinates.  The right panel removes the ambiguous regions, resulting in 4 track segments.  See text for  discussion on this type of common problem in SPT (i.e., process noise is commensurate with measurement noise).
\footnotesize 
 }
\label{fig:tracklets}
\end{figure}

Figure \ref{fig:tracklets} plots two simulated sample paths.  The paths are both constructed by a single Brownian motion process.  However,  
the two paths displayed are constructed using increments that are identical in magnitude, but opposite in sign (this is a demonstration of the reflection principle discussed in Ref.
\cite{billingsley}).  Around time 25, the two discretely sampled paths cross;  
the left panel  plots the truth (measurements from the two separate objects are denoted by different symbols).
Recall, that the observations are assumed to be point objects (the ``symbol attributes'' shown in the figure 
are not available to the researcher), therefore it is unclear which observations associate to which physical object. 
Furthermore, the researcher does not typically have accurate \emph{a priori} knowledge of the data generating process. 
  So in the toy illustrative example described here, there is not adequate statistical evidence to 
  confidently determine which measurements should be associated with pre-existing tracks.  
  Situations encountered in practical applications encounter more complex track ambiguity (as discussed later).
 In the right panel,
unambiguous segments are denoted by solid lines (these segments can only be determined when the position measurements sufficiently separate).  

The frame-to-frame association phase (identical to the track-to-measurement phase described in Sec. \ref{sec:approach}) of the original u-track algorithm \cite{Jaqaman2008} would force an association in the scenario shown in the left panel since measurements
are close to nominal tracks (note: the wrong association would be made). 
 However, an arc ambiguity or more advanced tracking algorithm leveraging information in multiple frames \cite{BP99,Poore2006,Chenouard2009a} could be used to break (or at least defer) the ambiguous frame-to-frame association made near time 25 in the left panel.  In the second phase (the gap-closing step) of u-track
 \cite{Jaqaman2008}, one attempts to join short track segments that do not overlap in time. 
Although we formulated the classic track-to-measurement  2D assignment problem  in Sec. \ref{sec:approach}, 
one can also consider linking measurement-to-measurement or track-segment-to-track-segment  spanning different times; the likelihood computation changes slightly in these scenarios \cite{BP99,Jaqaman2008} 
(in the GC phase, one merely needs to interpret a ``particle measurement'' as a 
track segment) 
 \footnote{Merging and splitting events are common in classic  \cite{BP99} and SPT tracking \cite{Jaqaman2008}; in such cases, 
 a merged composite track can contain measurements from multiple physical objects.  The UQ methods are readily applicable 
 to 2D merging and splitting events, but to facilitate exposition we omit this discussion in the main text and retain our 
 definition of a track (i.e., a collection of measurements believed to correspond to one physical object.)}.
The goal of GC is to produce  tracks containing measurements spanning a longer time.  In u-track, GC  associations are also formulated as a 2D assignment problem \cite{Jaqaman2008}, 
so  $k-$best assignment solvers can readily  be utilized. 
To construct the cost matrix of the GC step,  end points of short track segments are paired with the beginnings of other track segments (feasible solutions prevent temporal overlap;  an example of a cost matrix is shown in Fig. \ref{fig:QDbig}).

The problem illustrated in Fig. \ref{fig:tracklets} is that after the crossing event, there is no statistical evidence for stitching segments together even if one has exact knowledge of the data generating process.  For this example, looking at multiple frames of data \cite{BP99,Genovesio2006,Jaqaman2008,Chenouard2009a} via more sophisticated MHT approximations does not help in forming reliable tracks 
due to the statistical nature of discretely sampled systems where diffusive noise is a major contribution to the signal \cite{billingsley}. 
To make association decisions, standard
tracking algorithms assume an SDE or a collection of SDE models characterizing all possible motion types that can be encountered \cite{BP99}.   
The assumed models may be grossly inconsistent with reality; the real danger is that the \emph{a priori} models can heavily influence formed tracks and ``silently" introduce systematic biases. For example, 
if some of the models considered encourage straight line motion (e.g., SDEs with a constant drift driven by a standard Brownian motion are commonly used in SPT applications \cite{Genovesio2006}), then eventually tracks adhering to this enforced structure will be formed despite the lack of evidence for this motion in the raw data (in Fig. \ref{fig:tracklets}, for times 10-40, 
  an MHT using multiple frame tracking algorithm \cite{Chenouard2009a} considering a directed diffusion model in its cost matrix formulation may make grossly incorrect tracks after it observes enough of a trend it hopes to see).  

  The type of self-fulfilling prophecy described above introduces dangerous bias in scientific endeavors aiming to objectively  characterize molecular motion in cells.  In such applications, 
particles evolve in a complex, stochastic, time changing cellular environment. 
Each particle likely experiences different forces, friction, and thermal fluctuations (accurately quantifying these forces is difficult to accurately predict \emph{a priori} \cite{SPAdsDNA,SPAfric,pudi}). 
Crucial parameters, such as diffusion coefficients and measurement noise magnitude, can differ dramatically between tracks (and/or over time in a single molecule), but many cost functions assume that these parameters are frozen and equally applicable to all particles under consideration.  Some trackers attempt to estimate measurement noise on-the-fly \cite{Serge2008}, but parameters governing particle motion are difficult to reliably extract given only measurement data (where multiple associations can occur).
%
  If one has all of the information required to construct cost matrices, such as those formed using particle filters \cite{Arulampalam2002,Godinez2009}, then that information can be utilized (this might be the case when one is studying diffusion in a homogeneous medium).  However,  in live cell studies, one rarely has reliable knowledge of the functional form of the stochastic models governing the particle dynamics (even if one is fortunate enough to have access to a parametric form governing all particles, it is unlikely one  has accurate and unbiased \emph{a priori} knowledge on the parameter distribution defining the filters for \emph{in vivo} SPT studies).  For scientific applications aiming to extract unambiguous, objective, accurate models from experimental data, 
  we advocate the view that one should assume relatively coarse dynamical models imposing minimal motion assumptions (such as pure diffusion with a relatively large diffusion coefficient)  
  in track formation.   The SPT algorithms demonstrated \cite{Serge2008,Jaqaman2008} readily
  allow tracking with minimal modeling assumptions.

\subsection{Extending ``u-track" to  Include Association UQ}
\label{sec:GCUQ}

  Next, the Correspondence Method  is  used in conjunction with the algorithm proposed in Ref. \cite{Jaqaman2008}.  Our algorithm can be used  to aid  both association phases of the u-track.  For the first frame-to-frame association phase, one can set a threshold, $P_{F2F}$.    
 Another threshold, $P_{GC}$, can be set for stitching segments together taking place in the second gap-closing phase.  In the former, any association in the  optimal solution $k=1$ having an arc ambiguity below $P_{F2F}$ can be revoked (forcing track initiation).  Similarly, in the GC phase, arc associations below $P_{GC}$ can be rejected (this prevents questionable track stitching).  The  result is  a collection of track segments where one is fairly confident in the underlying arc associations.  By rejecting gap-closing associations,
one may produce two tracks generated from one underlying molecule, but $P_{GC}$, can be adjusted to balance track segment extension
vs. association ambiguity.


\subsubsection{Application to Extracting Tracks in Crowded Cell Plasma Membrane Environments}
To illustrate the u-track modifications discussed previously, we analyze the motion of quantum dot (QD)-labeled voltage gated potassium channels lacking the last 318 amino acids of the C-terminus,  $\Delta$C318 Kv2.1 \cite{Tamkun2007}. The truncation of the C-terminus has been shown to allow Kv2.1 to diffuse freely on the plasma membrane \cite{Tamkun2007} whereas wild-type Kv2.1 displays confined motion \cite{Tamkun2007,Weigel2011}. The desired dense labeling in combination with the inherent blinking behavior QDs, presents serious challenges to current SPT tracking algorithms. This thereby creates an ideal system in which to test our algorithm. By pruning questionable tracks, our approach can aid in the  analysis of ion channel motion in a complex biological system.

 \subsubsection{Experimental Setup}

Human embryonic kidney (HEK) 293 cells were induced to express 3 $\mu$g/$\mu$L of the mutant channel $\Delta$C318 Kv2.1 as described previously \cite{Tamkun2007}. The construct contains an extracellular biotin accepting domain allowing the channel to be enzymatically biotinylated and labeled with a streptavidin conjugated QD (Qdot-655, Invitrogen). Cells were imaged in a home-built, objective-type total internal fluorescence microscope, under excitation with a 1 mW 473 nm laser line \cite{Weigel2011,Weigel2012}. Fluorescence was captured in a cooled back-illuminated electron-multiplied charge couple device (EMCCD, Andor). Movies were acquired using Andor IQ software at a frame rate of 20 frames per second. 

\begin{figure}[htb]
\begin{minipage}[l]{.575\linewidth}
  \centering
\def\pw{.99}
  \centerline{\includegraphics[width=.99\textwidth]{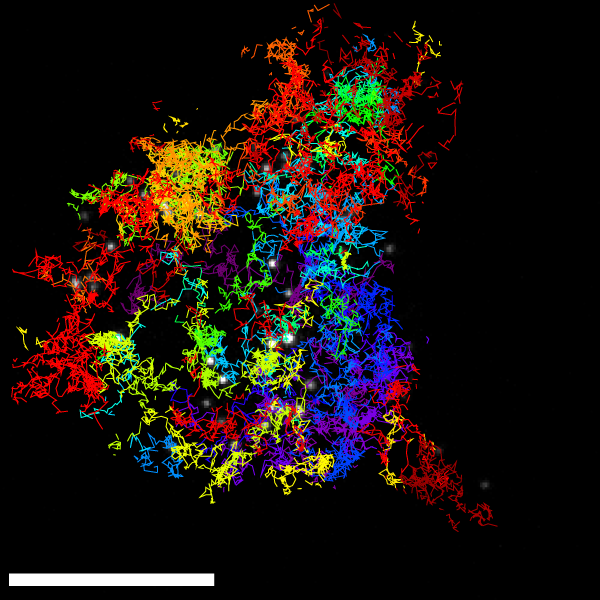}}
   \centerline{\footnotesize 
   Ion channel tracks.  Scale-bar denotes $10 \mu m$ }
\end{minipage} 
\begin{minipage}[r]{.4\linewidth}
  \centering
\def\pw{.99}
  \includegraphics[width=\pw\textwidth]{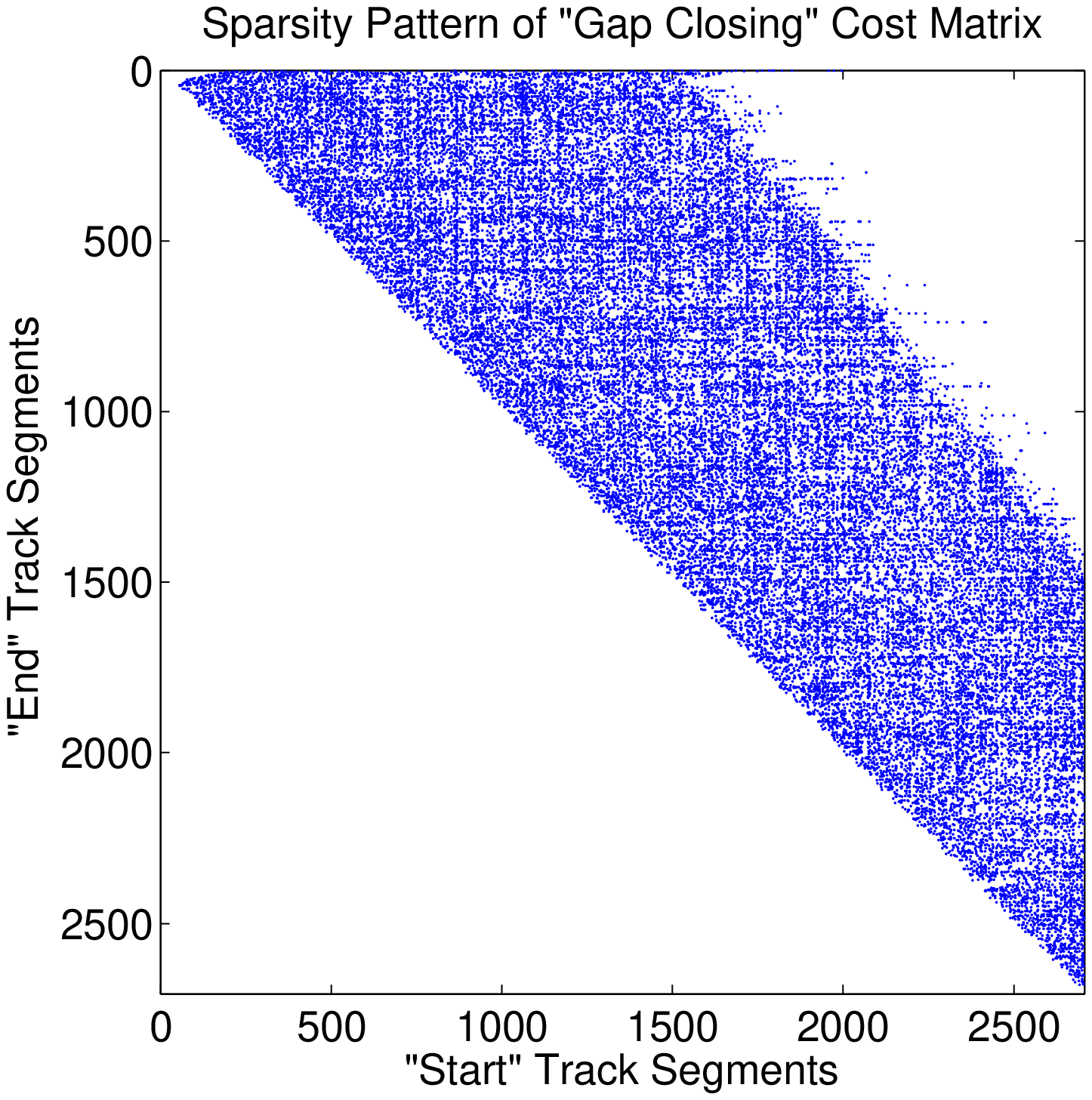} 
  \centerline{\footnotesize }
\includegraphics[width=\pw\textwidth]{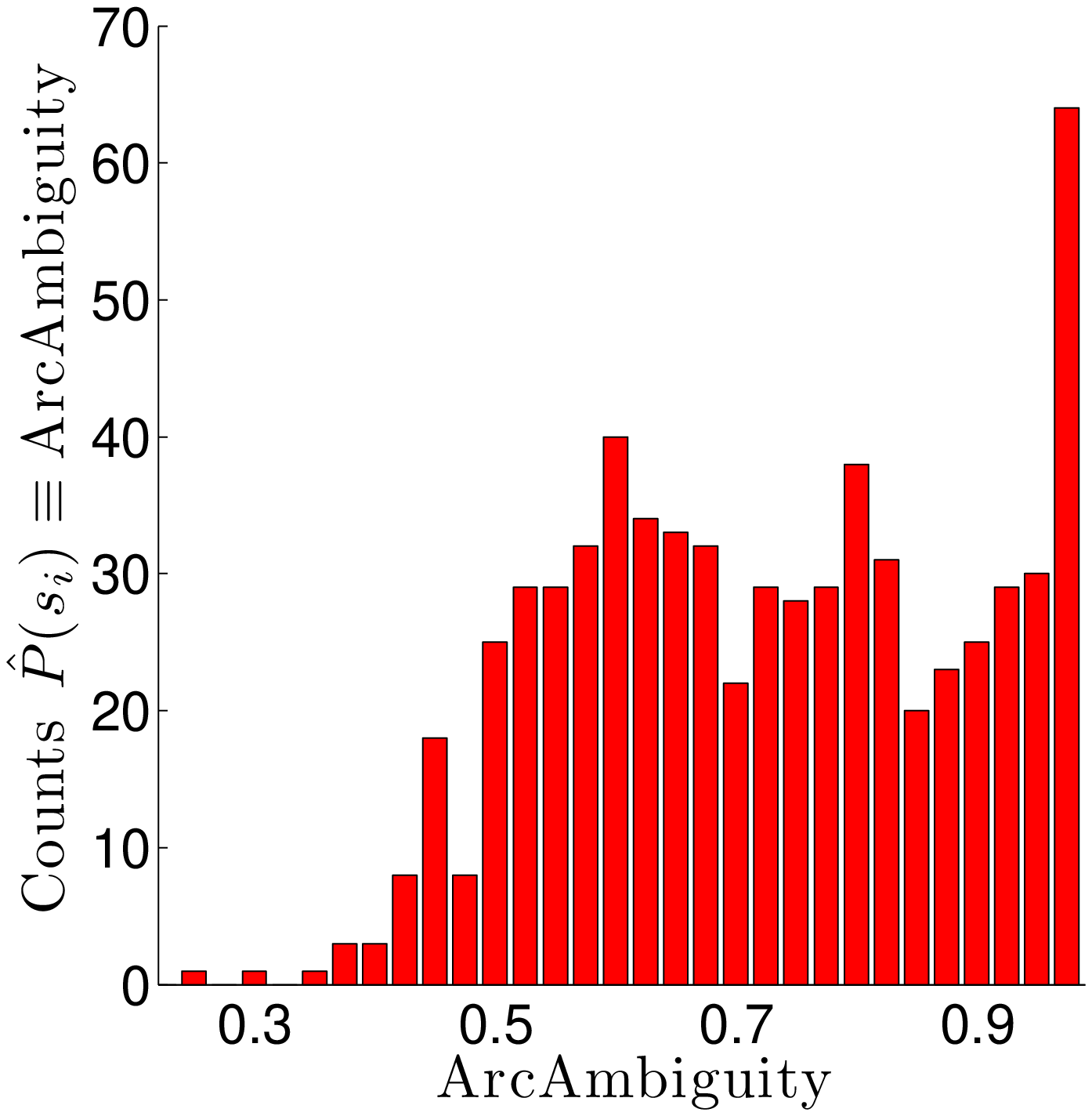}
   \centerline{\footnotesize Histogram of low probability GC arcs.}
\end{minipage}
\centerline{\footnotesize }
\caption{ 
\footnotesize 
The left panel displays a single frame of a microscope image where quantum dots are attached to a protein (the ion channel Kv2.1 \cite{Tamkun2007} with a mutation outlined in main text).  The
motion of multiple proteins was tracked in human embryonic kidney cells (the color overlay in the left panel displays some of the tracks generated by analyzing 400 frames of data with u-track \cite{Jaqaman2008}).
The top right panel displays the sparsity pattern of the ``gap-closing'' cost matrix employed by u-track (there are a total of 2705 track segments that can be connected after the 400 images were processed via the frame-to-frame phase of u-track).   The rows labeled ``End Track Segments'' denote track indices corresponding to tracks that terminated before the last frame; the columns labeled as ``Start Track Segments'' indicate track indices of tracks initiated after frame one. 
A non-zero entry in the cost matrix suggests a possible linking due to lack of temporal overlap and adequate spatial proximity of  the two segments at the start and end points.
The bottom right panel displays  the estimated arc probabilities in the optimal solution  falling below the $P_{GC}$ threshold (here $P_{GC}=0.99$).  For each arc in the optimal  solution, $\{ s: s\in H_1 \}$, we estimated $\hat{P}(s)$ and subsequently used the Correspondence Method and hypothesis $H_1$ to generate the $\hat{P}(s)$'s (the plot contains the histogram of $\{ \hat{P}(s): s\in H_1 \}$).  
 }
\label{fig:QDsnap}
\end{figure}


\begin{figure}[htb]
\center
\centering
\def\pw{.475}
\centerline{\footnotesize \textbf{ \Large Distribution of $\hat{P}(s_i)<0.99$ for $k=10$}}
\begin{minipage}[b]{.425\textwidth}
  \def\pw{.99}
  \centerline{\includegraphics[width=\pw\textwidth]{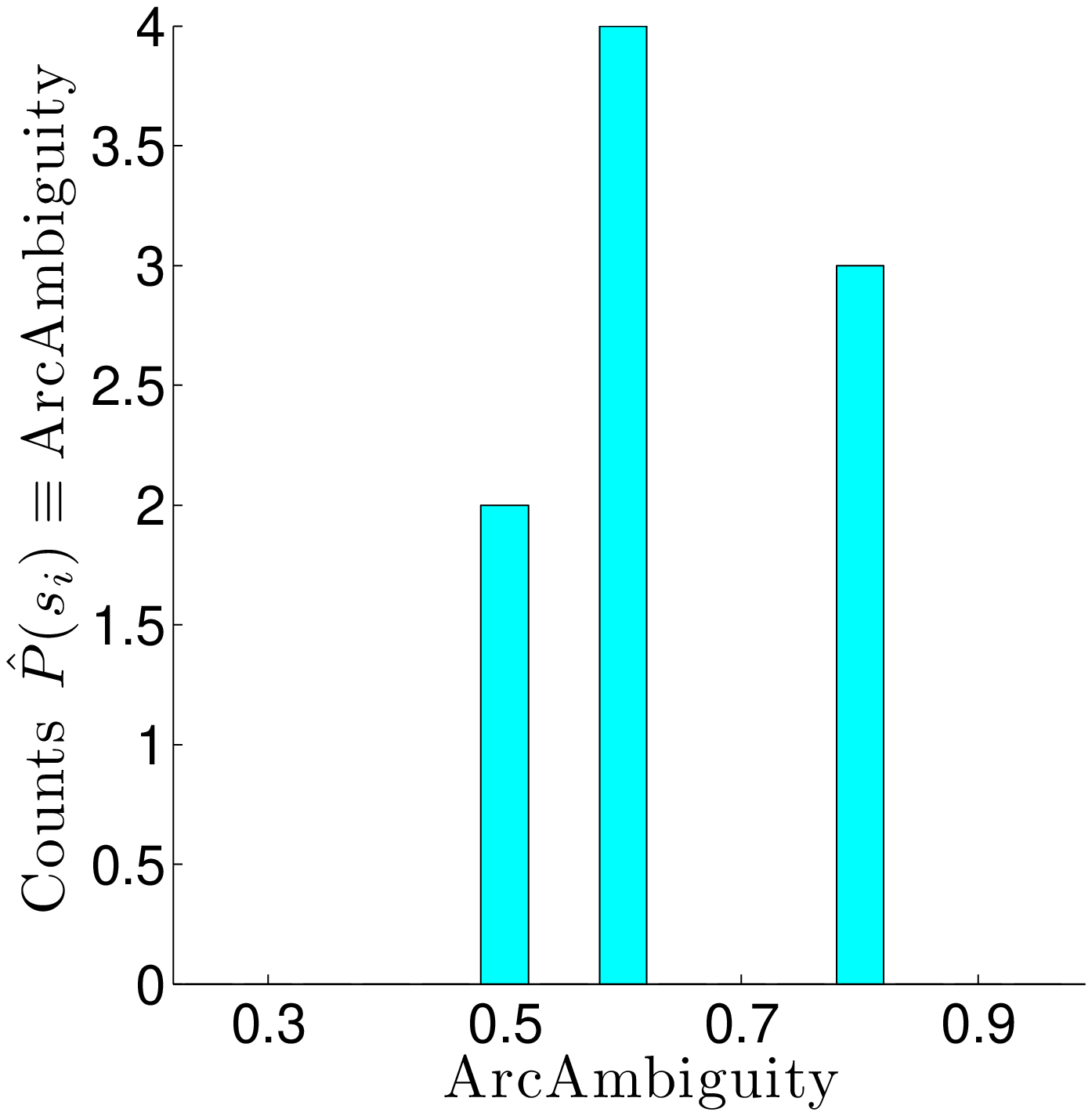} } 
   \centerline{\includegraphics[width=\pw\textwidth]{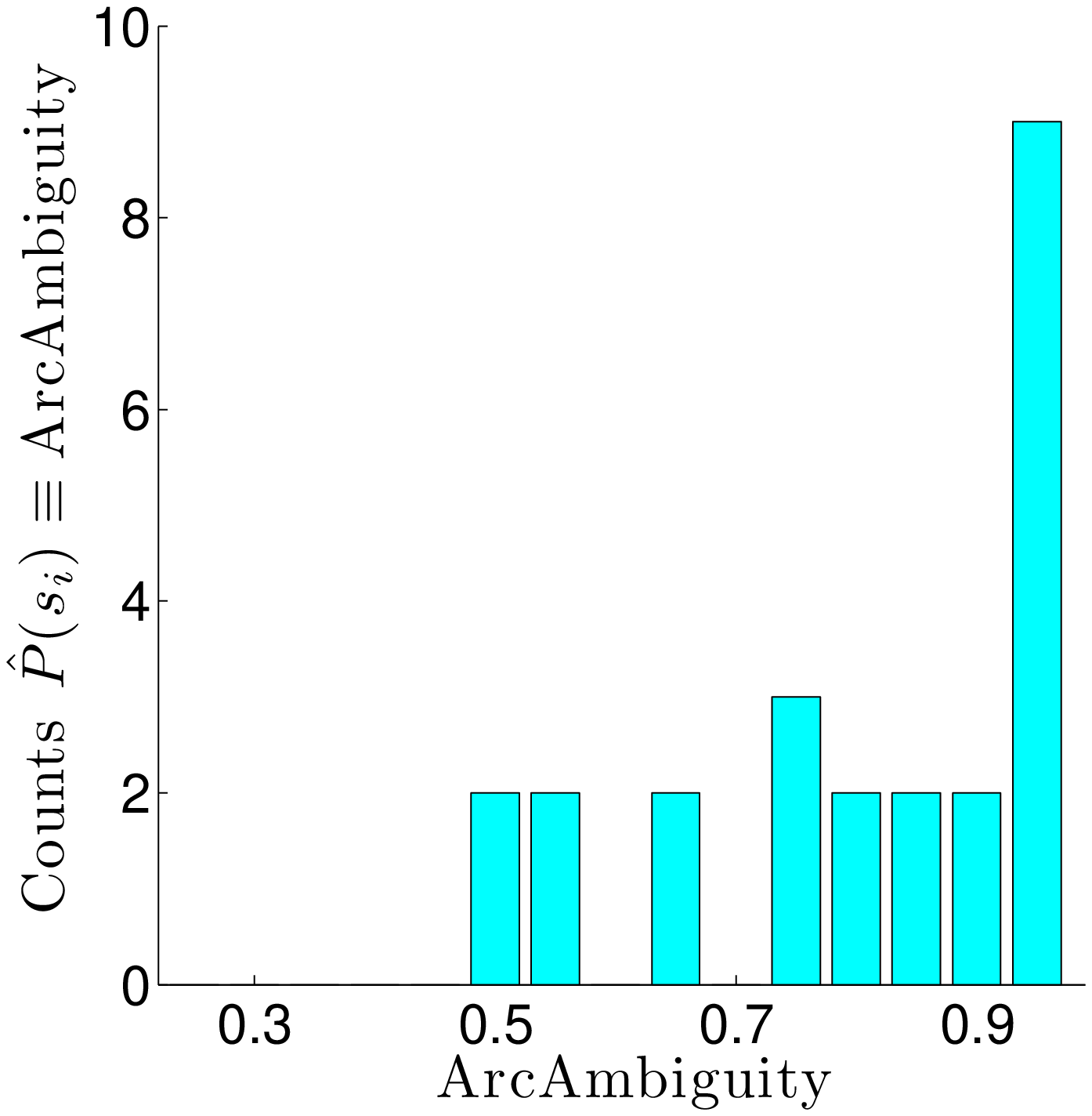} } 
 \end{minipage} 
 \begin{minipage}[b]{.425\textwidth}
  \def\pw{.99}
  \centerline{\includegraphics[width=\pw\textwidth]{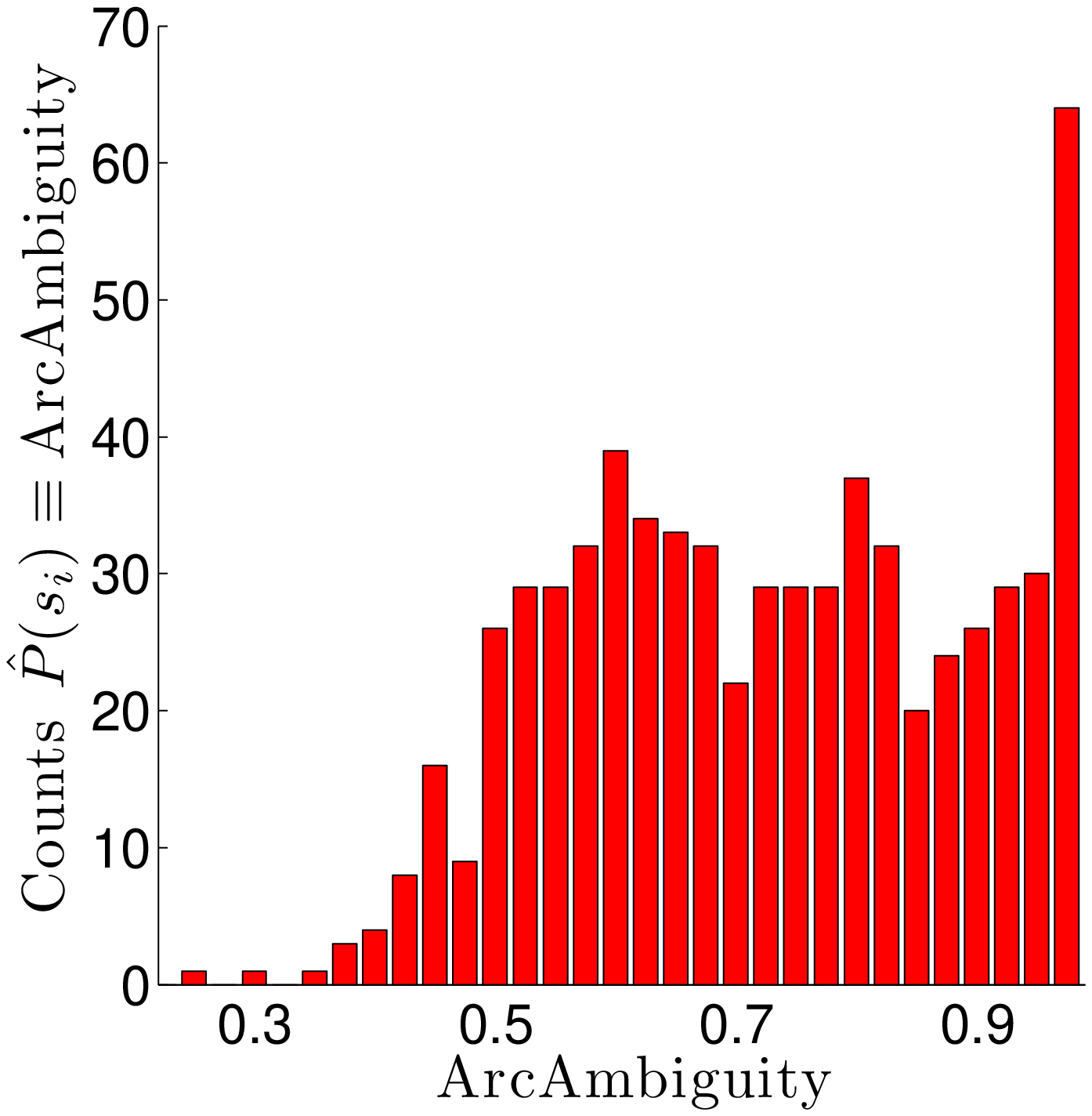}}
  \includegraphics[width=\pw\textwidth]{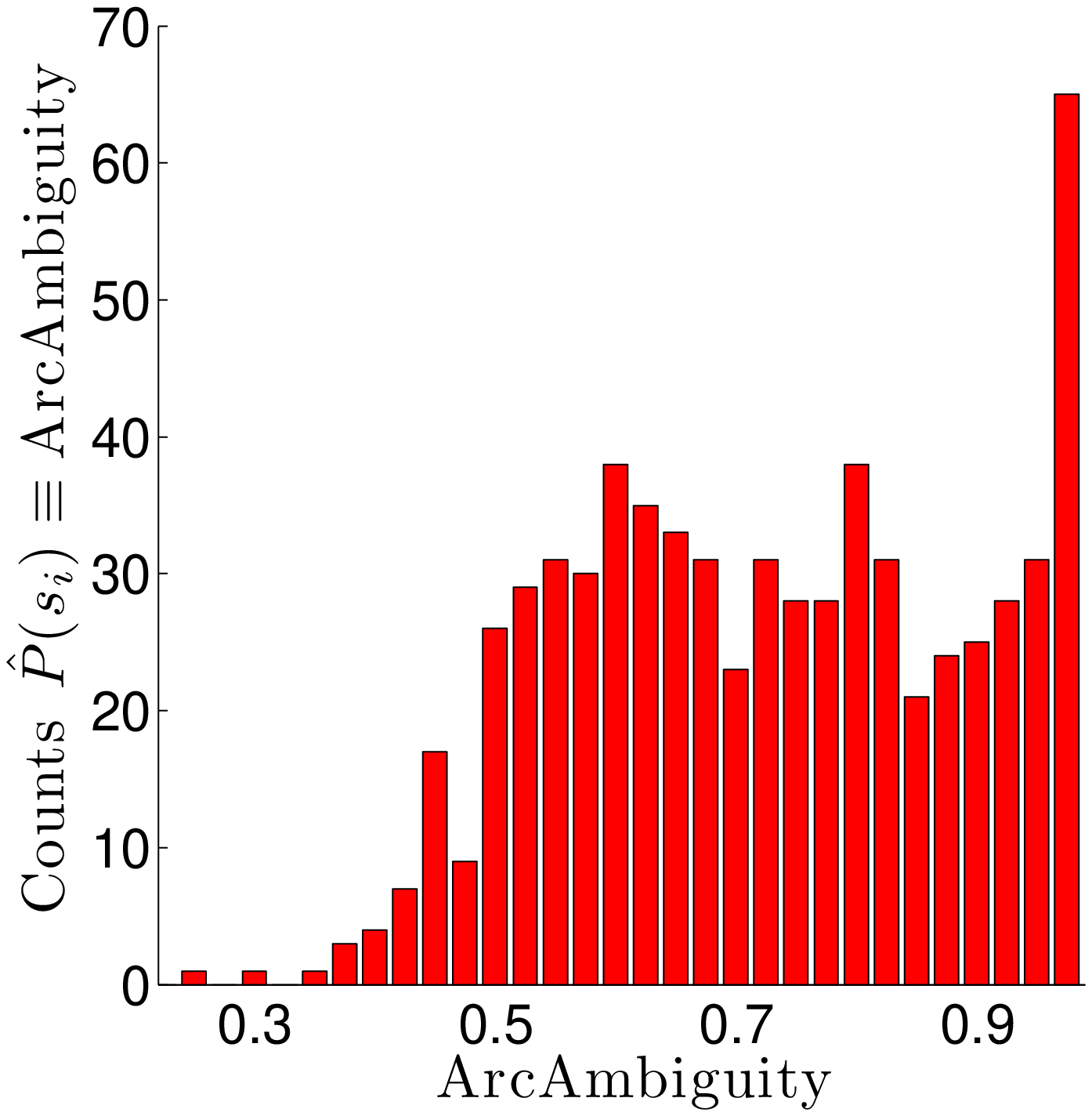}
  \end{minipage} 
   \centerline{\footnotesize \textbf{ \Large Distribution of $\hat{P}(s_i)<0.99$ for $k=100$}}
\centerline{\footnotesize }
\caption{ 
\footnotesize 
Histograms of the estimated arc probabilities in the optimal solution  falling below the $P_{GC}$ threshold (here $P_{GC}=0.99$).  For each arc $s_i$ in $H_1$ (subscript denotes optimal solution), we estimated $\hat{P}(s_i)$.
The $\hat{P}(s_i)$'s were computed using two values of $k$ (10 and 100) using both the Traditional Method (left panels) and the Correspondence Method (right panels). The $k=1$ case for the Correspondence Method is displayed in the bottom right panel of Fig. \ref{fig:QDsnap}; the distribution of the $\hat{P}(s_i)$'s changes in  a negligible fashion when $k$ is increased (careful inspection of the histograms shows minor differences).  The $k=1$  case is trivial for the Traditional Method (i.e., $\hat{P}(s_i)=1$ for all $i$) so it is omitted.
 }
\label{fig:QDbig}
\end{figure}

 A MATLAB-based tracker augmenting the publication in Ref. \cite{Jaqaman2008} was used to analyze a sequence of 400 images each receiving a camera exposure time of 50ms. Figure \ref{fig:QDsnap} displays some representative tracks generated in this experiment.  The parameters used in the tracking routine are reported in the Appendix. To illustrate the power and practical utility of the Correspondence Method,
 we replaced the JV assignment solver with a modified Murty algorithm capable of processing rectangular cost matrices.
 The Murty algorithm subsequently fed  the Traditional and Correspondence Method $k$-best solutions  \cite{drummond90,Kragel2012}. 
   In addition, the procedure discussed in Sec. \ref{sec:GCUQ} using $P_{GC}=P_{F2F}=0.99$ was applied.  With the Correspondence Method,  setting $P_{F2F}=0.99$ and $k=1$ resulted in 2658 segments being initiated after frame one and 2656 track segments terminating before the final frame in the frame-to-frame association phase of the modified u-track \cite{Jaqaman2008}. If no UQ pruning was applied, i.e. $P_{F2F}=0$, the tracking algorithm has only 2132 segments being initiated after the first frame and 2128 track segments terminating before the final frame; revoking questionable frame-to-frame associations clearly increases the number of track segments.  The second ``gap-closing'' phase attempts to put these segments back together.  A total of 2705 segments defined the square GC gap-closing cost matrix in the formulation of Ref. \cite{Jaqaman2008}.

   The sparsity pattern of the square cost matrix defining the resulting gap-closing assignment problem is 
   displayed in the upper right panel of Fig.  \ref{fig:QDsnap}. Non-zero entries in this cost matrix indicate  track segments that (i) do not share measurements overlapping in time and (ii) are spatially close enough (an input gating parameter \cite{Jaqaman2008})to be considered for GC track stitching. Note how the diagonal entries of the gap-closing cost matrix are all zero since a track segment temporally overlaps with itself (other zero entries are a result of either spatial or temporal gating).
    Only 389 out of the 2705 track segments were unassigned in the $k=1$ solution.  In this crowded dense scenario with high diffusivity, this suggests that many track segments often cross and can be assigned with the default parameters \cite{Jaqaman2008} defining the gap-closing cost matrix. 
      The bottom right panel of Fig.  \ref{fig:QDsnap} displays a histogram of the $\hat{P}(s_i)$ for non-null arcs $s_i \in H_1$ whose ambiguities 
   satisfy $\hat{P}(s_i)<P_{GC}=0.99$   computed using the Correspondence Method and $\HypsSample=H_1$.  Of the non-null arcs in $H_1$,
   $28.7\%$ are below  $P_{GC}$ suggesting that a substantial fraction of track segments stitched together via GC are questionable.   Note that for $k=1$, the modified Murty assignment solver and the original JV solver used in Ref. \cite{Jaqaman2008} provide the same association output.
   %
   It should also be mentioned, given set of tracking parameters (often containing many tunable gating parameters \cite{BP99,Jaqaman2008,Serge2008}), that the Correspondence Method can be used to quantify the quality of tracks formed for a given population of particles (the quality of tracks depends on many factors, e.g. density, diffusivity of tagged particles, and measurement noise).

   Figure  \ref{fig:QDbig}
   displays the distribution of $\hat{P}(s_i)$,  using $k=10$ and $k=100$ obtained using both the Traditional and Correspondence Method.   The Correspondence Method's distribution of $\hat{P}(s_i)$'s is relatively independent of $k$ (the panels in Fig.  \ref{fig:QDbig} complement the bottom right panel of Fig. \ref{fig:QDsnap}).  Using $k=100$ and the Correspondence Method, $28.9\%$ of the non-null arcs in the optimal solution fall below the $P_{GC}$ threshold. 
    Insensitivity of $\hat{P}(s_i)$ to $k$
   confirms that a significant fraction of tracks generated in the second phase of this routine  are of dubious quality due to ambiguous  gap-closing assignments.    

    The Traditional method has poor UQ estimates in this large problem for $k \in [1,100]$ because many hypotheses have similar scores. 
    There are small numerical differences between the top scoring hypotheses; furthermore, these high scoring hypotheses  only differ by a few arcs.  To make things concrete and to highlight why when using the Traditional Method to compute arc ambiguities so few $\hat{P}(s_i)$'s fall below $P_{GC}$,  
     the set difference between the top scoring hypothesis and the four other top hypotheses are provided:
    $H_1-H2=\{s_a,s_b\}$, $H_1-H3=\{s_c,s_d\}$, and $H_1-H4=\{s_a,s_b,s_c,s_d\}$ (the actual arc numbers are uninformative).  In the top solutions, a permutation of a small number of arcs are going in and out of the top  hypotheses.    The likelihood ratio of $L(H_{100})=0.9513 \times L(H_{1})$. This is suggestive that the field of view of the microscope is too crowded using the tracking parameters and algorithm employed.   The Traditional Method slowly explores the phase space of all feasible hypotheses, hence using $k=100$, produces a poor estimate of the arc ambiguities.

    The Correspondence Method quickly explores the space of feasible arcs since
    arcs of interest can be systematically  broken. The cluster probability computation permits one to approximate
    the relative importance of that arc. 
     If one has the computational resources and time to set $k$ very large, then the Traditional Method's distribution would approach that of the Correspondence Method.  However, for this large system, with $k=100$ we pushed the RAM limits of the 8GB computer utilized in computations; with the $k$-best solver employed (finding the $k=100$ best solutions on a laptop took under 4 minutes, and the post-processing of the data
using the Correspondence Method took under 20 seconds).  More sophisticated $k-$best routines beyond the modified Murty algorithm can allow for larger $k$ values in this problem with the same hardware, but  
   a major advantage of the Correspondence Method is that one does not need to use excessively large $k$ values to obtain reasonable UQ accuracy.  We demonstrated that the  problem of selecting a ``large enough" value for $k$ (which plagues the Traditional Method \cite{Kragel2012}) can be substantially mitigated if one uses the Correspondence Method. Since $k=1$ results are meaningful, one can use off-the-shelf linear assignment solvers and the Correspondence Method to assess the quality of tracks generated (this feature is practically attractive since not all researchers have easy access to state-of-the-art discrete optimization routines).

\section{Conclusions}

A new algorithm for processing the uncertainty in data association has been introduced.  The technique forms new feasible hypotheses from given  solutions to determine the ambiguity (i.e, discrete data association UQ) in  arcs of interest.  A procedure for fusing discrete optimization output (provided via any type of assignment solver) with inferred information to improve arc ambiguity accuracy was demonstrated on small scale control examples as well as large-scale problems.  The basic intuition behind the method was presented via a variety of examples. Practical computational implementation details and a complexity analysis required for better understanding large-scale performance is provided  in the Appendix.

The method shows great promise for quantifying uncertainty when there are multiple hypotheses having scores that are not well separated (i.e., a single global optimal hypothesis does not dominate given the metrics defining the cost matrix). The method is fast, and only introduces marginal computational effort in real-time applications.  In  large-scale problems using off-line computations, memory may be exhausted before a good representative sample of association hypothesis can be constructed (hence the accuracy of traditional UQ methods can suffer greatly \cite{Kragel2012});  in such situations, we demonstrated that the Correspondence Method can help in providing more reliable association UQ estimates.  
The algorithm can also readily wrap around various trackers to generate UQ information.   In applications where the underlying solver may be hard to access or modify, the Correspondence Method can still produce reasonable UQ estimates provided only a single association hypothesis (though accuracy improves as more hypotheses are provided). There are numerous application domains where such technology can be utilized.

 Regarding the SPT biophysics and cell biology applications focused on this paper, standard tracking algorithms  in this problem domain \cite{Genovesio2006,Jaqaman2008,Serge2008,Danuser2011,Meijering2012} typically input heuristic dynamical models (making many questionable assumptions) for creating metrics used to define various data association cost matrices. The Correspondence Method can  aid in screening the vast number of tracks and determine which ones should be subjected to more demanding computational processing aiming to more accurately fit and test stochastic models characterizing the experimental data. 
 If researchers can quantify probabilities of more complex physical events  
 (e.g., endocytosis or molecular binding), our method  can also be used to identify when these events of biological relevance can possibly describe the sequence of images generated by the  microscope.   
 However, in the biological SPT problem domain, we advocate minimizing  \emph{a priori} assumptions about expected particle motion.  That is, posit simple stochastic rules that can cover all types of motion that can be encountered
 instead of imposing complex models to make initial associations that may introduce systematic bias in the data association phase.  Once unambiguous tracks (or track segments) are in hand, one can consider other time series modeling techniques \cite{SPAdsDNA,SPAfilter} to extract detailed molecular kinetic information from the collection of reliable tracks (e.g., using methods such as those discussed in Ref. \cite{SPAgof}).  In addition, one can consider other stitching mechanisms making better use of kinetic information inferred from  the time series of unambiguous track segments. 
  Regardless if one agrees with our viewpoint, the discrete optimization track UQ tools we introduce can be utilized to  quantify the ambiguity in  tracks formed via 2D association solvers. If one prefers using questionable sophisticated models of motion in the early stage of tracking (hence potentially biasing results), our approach can still be leveraged to quantify when scenarios become too dense to form reliable tracks conditioned on one's prior assumptions.

{
\section*{Acknowledgments}
\noindent We thank Scott Lundberg (Numerica), Ben Slocumb (Numerica), and Lucien Weiss (Moerner Lab at Stanford U.) for helpful comments on an earlier draft of this manuscript.  
 We are also a greatful to Michael Tamkun (CSU) and Liz Akin (CSU) for their help with the Kv2.1 mutant. 
 DK was financially supported by the National Science Foundation under Grant 0956714.  CPC was financially supported by internal research and development funds from Numerica Corporation.
 \\
}


\appendix
 \section{Illustration of Cluster Formation Procedure}

\begin{figure}[htb]
\center
\begin{minipage}[b]{.33\linewidth}
  \centering
\def\pw{.99}
  \centerline{\includegraphics[width=\pw\textwidth]{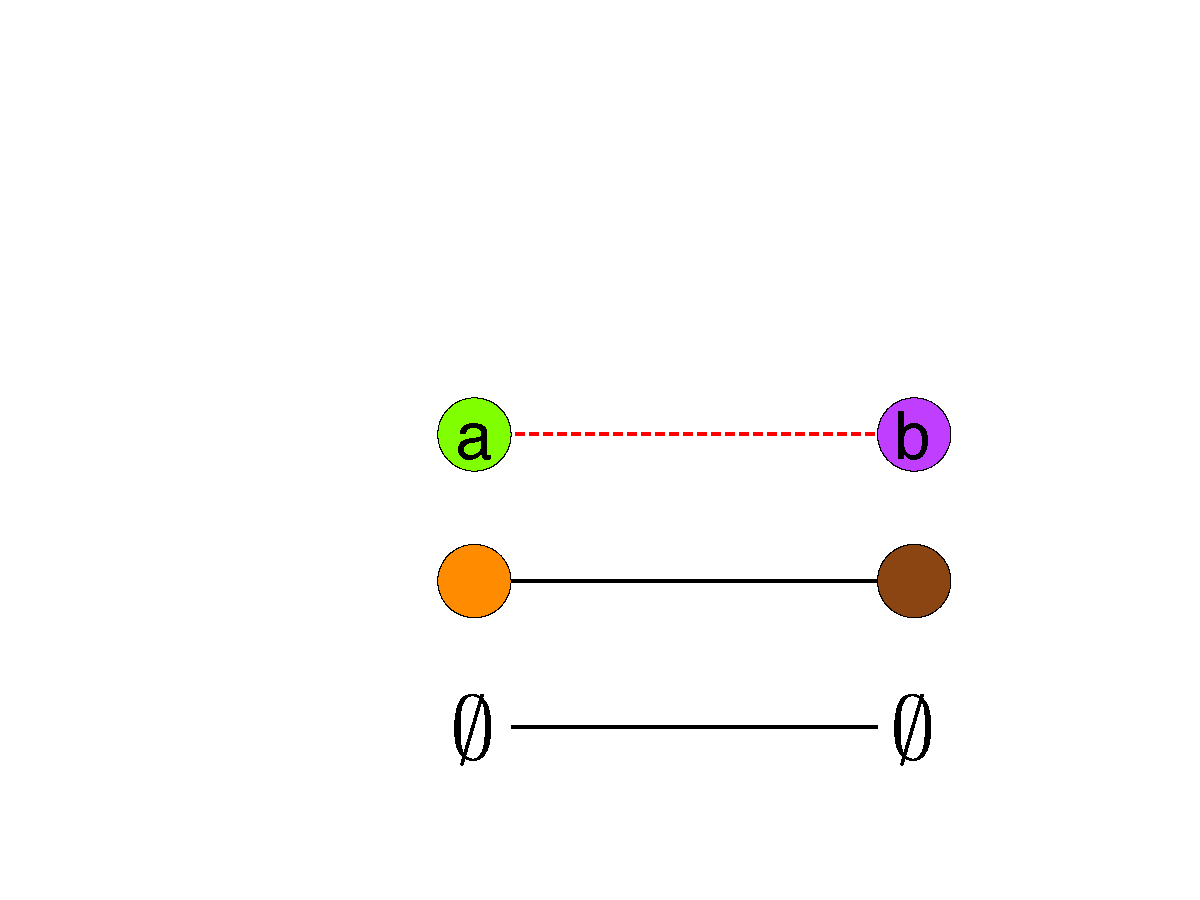}}
   \centerline{\footnotesize \textbf{Defining hypothesis in cluster $C^{ab}(H_1)=C^{s_1}(H_1)$ (target arc  denoted by red-dashed line).} }
\end{minipage}

\begin{minipage}[b]{.33\linewidth}
  \centering
    \def\pw{.99}
    \centerline{\includegraphics[width=\pw\textwidth]{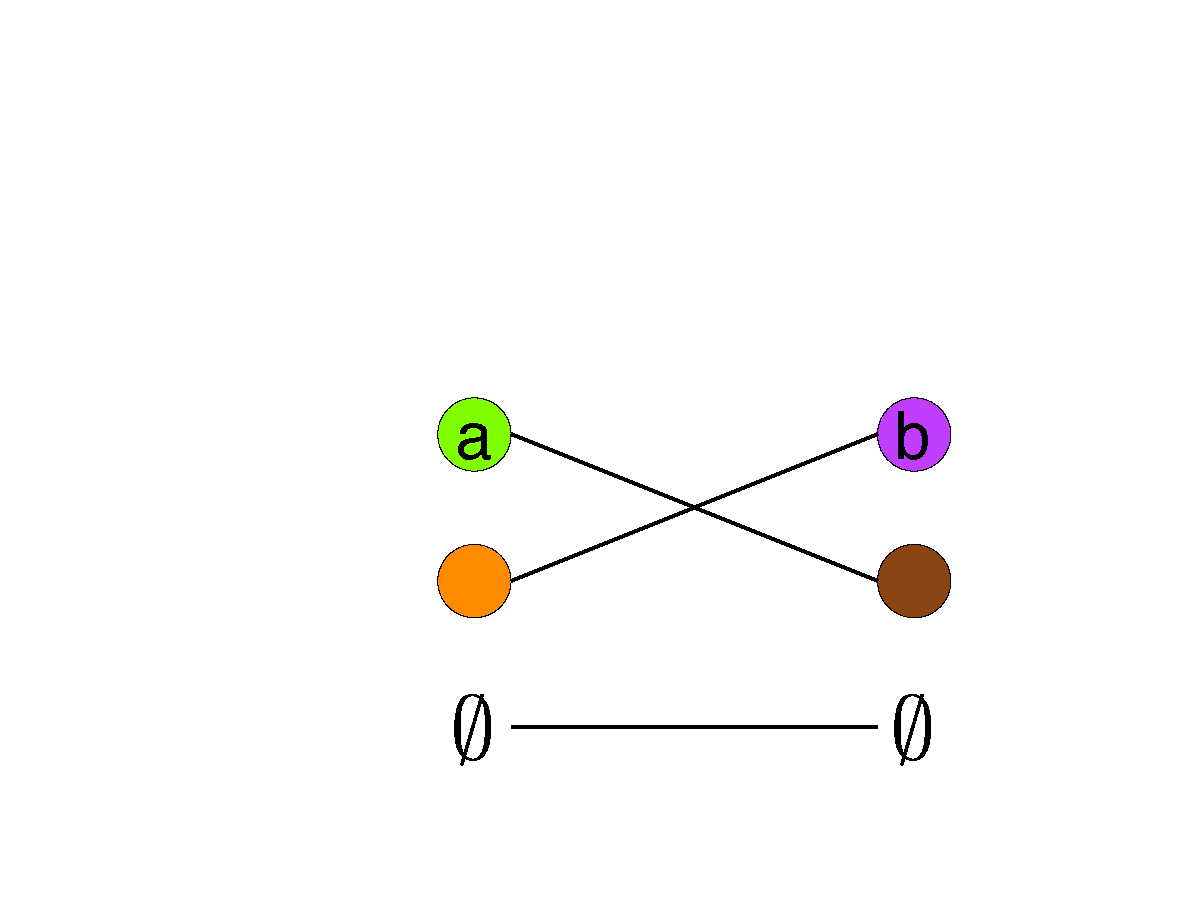}}
  \centerline{\footnotesize  \textbf{Hypothesis (ii) in cluster $C^{s_1}(H_1)$} }\medskip
\end{minipage}
\hfill
\begin{minipage}[b]{0.33\linewidth}
  \centering
   \def\pw{.99}
    \centerline{\includegraphics[width=\pw\textwidth]{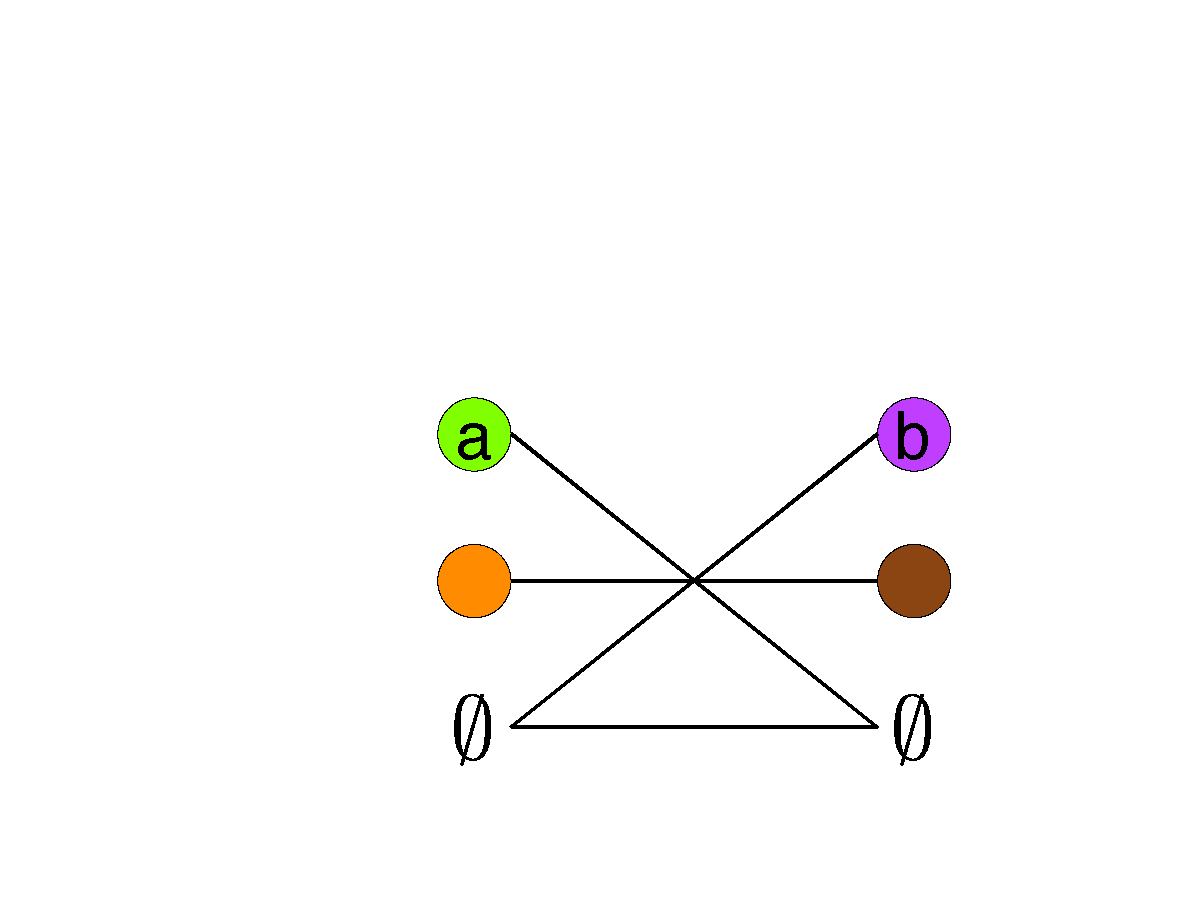}}
  \centerline{\footnotesize  \textbf{Hypothesis (iii) in cluster $C^{s_1}(H_1)$} }\medskip
\end{minipage}

\caption{
\footnotesize 
 Illustration of forming clusters defined by arc ab.  The target arc shown
is contained in  the $k=1$ hypothesis (i.e., the global optimum). 
 In this simple $2 \times 2$ example, the hypotheses above  make up one of two possible clusters.  The hypotheses in each cluster are formed by swapping the target arc with the other arcs in hypotheses containing the arc $ab$ (this arc only appears in two hypothesis in this $2\times 2$ problem).  
 The hypotheses above making up one cluster which is denoted by $C^{ab}(H_1)$.
  Switching the target arc with the other nonzero arc in the problem generates the bottom
left hypothesis in cluster $C^{ab}(H_1)$; switching with  $(\emptyset, \emptyset)$ generates the bottom right hypothesis.  
 }
\label{fig:res2a}
\end{figure}

\begin{figure}[htb]
\center
\begin{minipage}[b]{.4\linewidth}
  \centering
\def\pw{.99}
  \centerline{\includegraphics[width=\pw\textwidth]{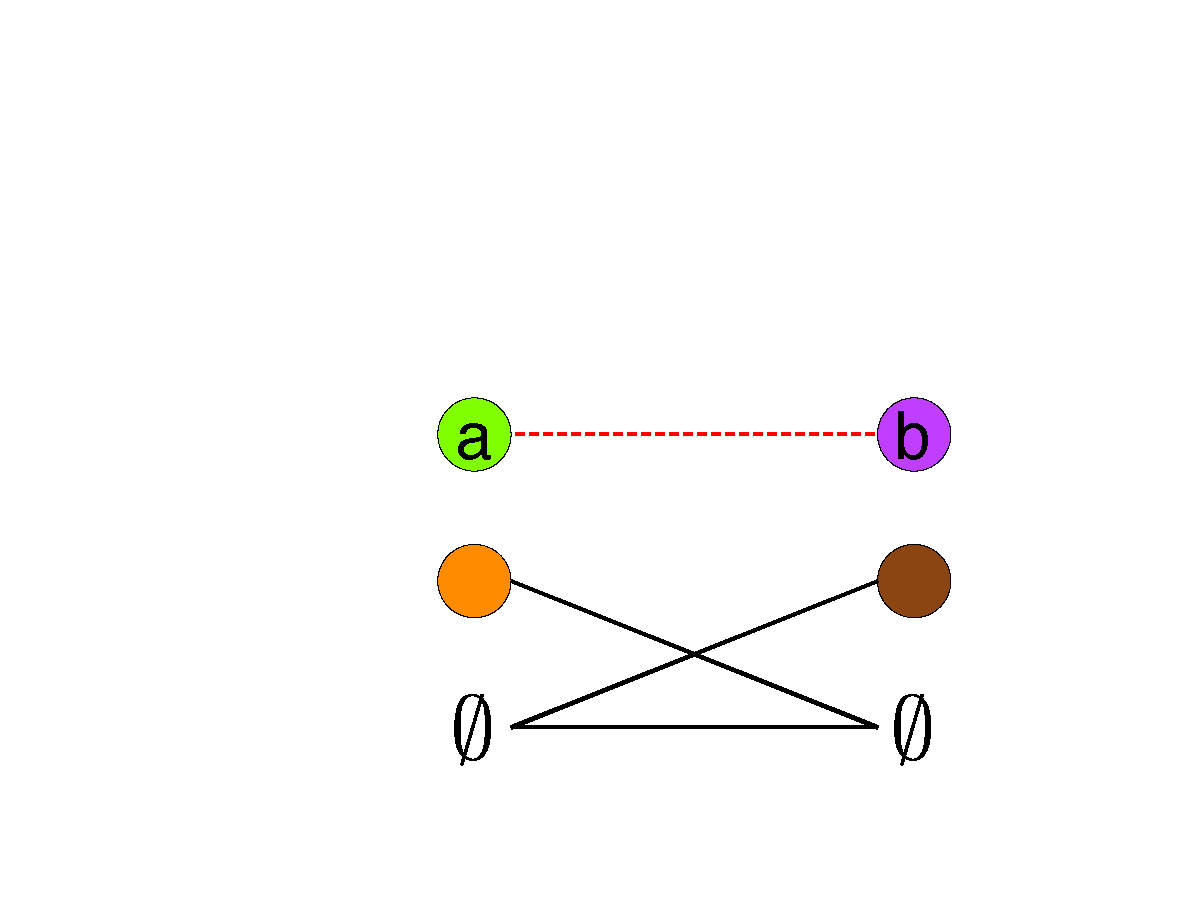}}
  \centerline{\footnotesize \textbf{Defining hypothesis in cluster $C^{ab}(H_3)=C^{s_1}(H_3)$ (target arc denoted by red-dashed line).}}
\end{minipage}

\def\pw{.4}
\begin{minipage}[b]{\pw\linewidth}
  \centering
    \def\pw{.99}
    \centerline{\includegraphics[width=\pw\textwidth]{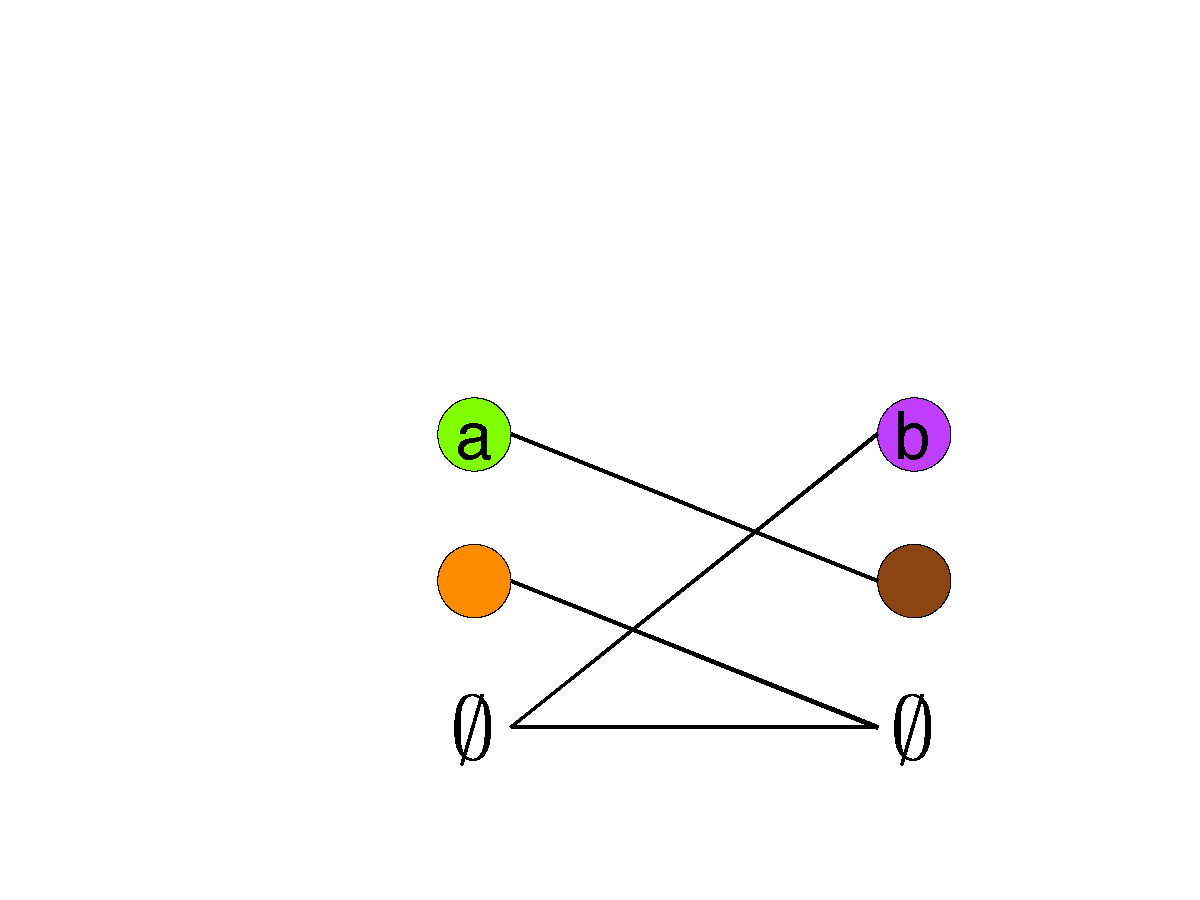}}
  \centerline{\footnotesize  \textbf{Hypothesis (ii) in cluster $C^{s_1}(H_3)$} }\medskip
\end{minipage}
\begin{minipage}[b]{\pw\linewidth}
  \centering
   \def\pw{.99}
    \centerline{\includegraphics[width=\pw\textwidth]{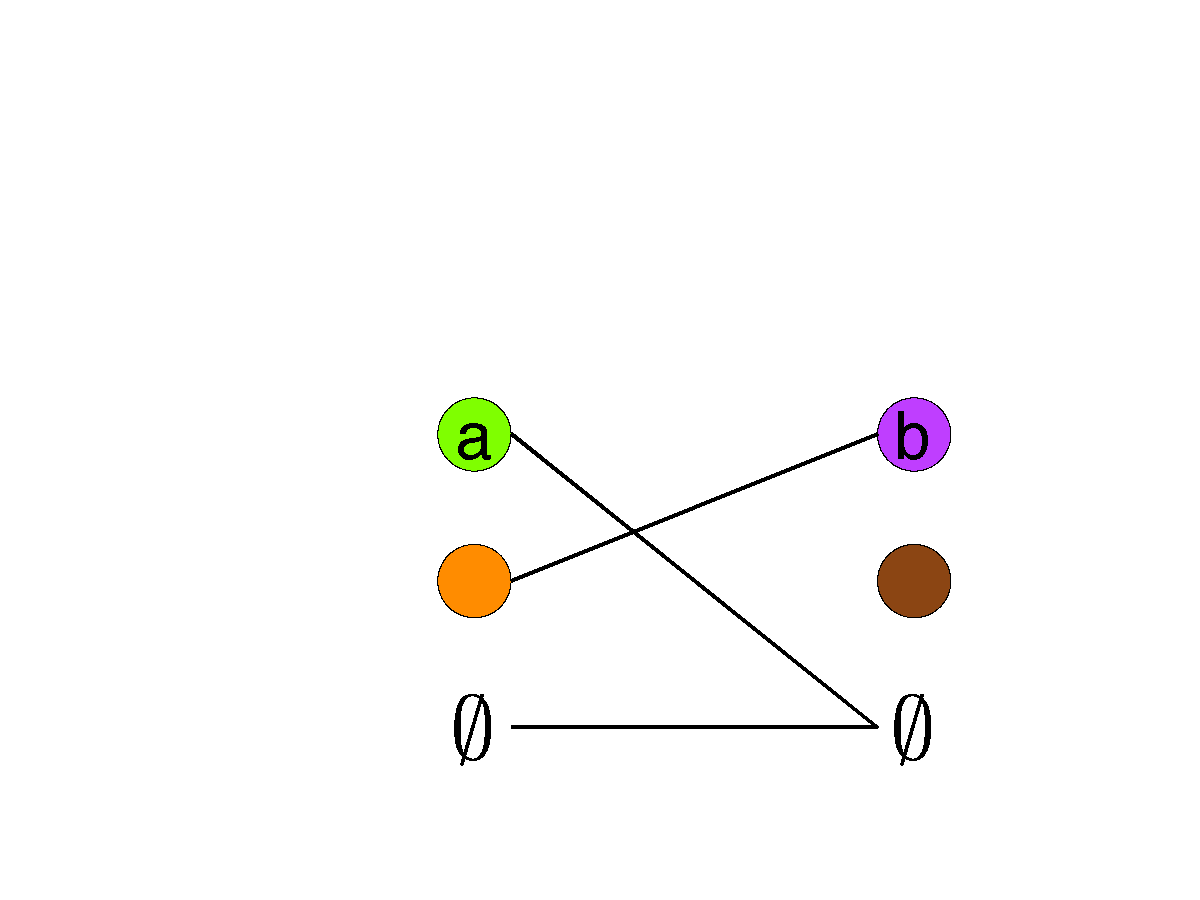}}
  \centerline{\footnotesize  \textbf{Hypothesis (iii) in cluster $C^{s_1}(H_3)$} }\medskip
\end{minipage}
\begin{minipage}[b]{\pw\linewidth}
  \centering
   \def\pw{.99}
    \centerline{\includegraphics[width=\pw\textwidth]{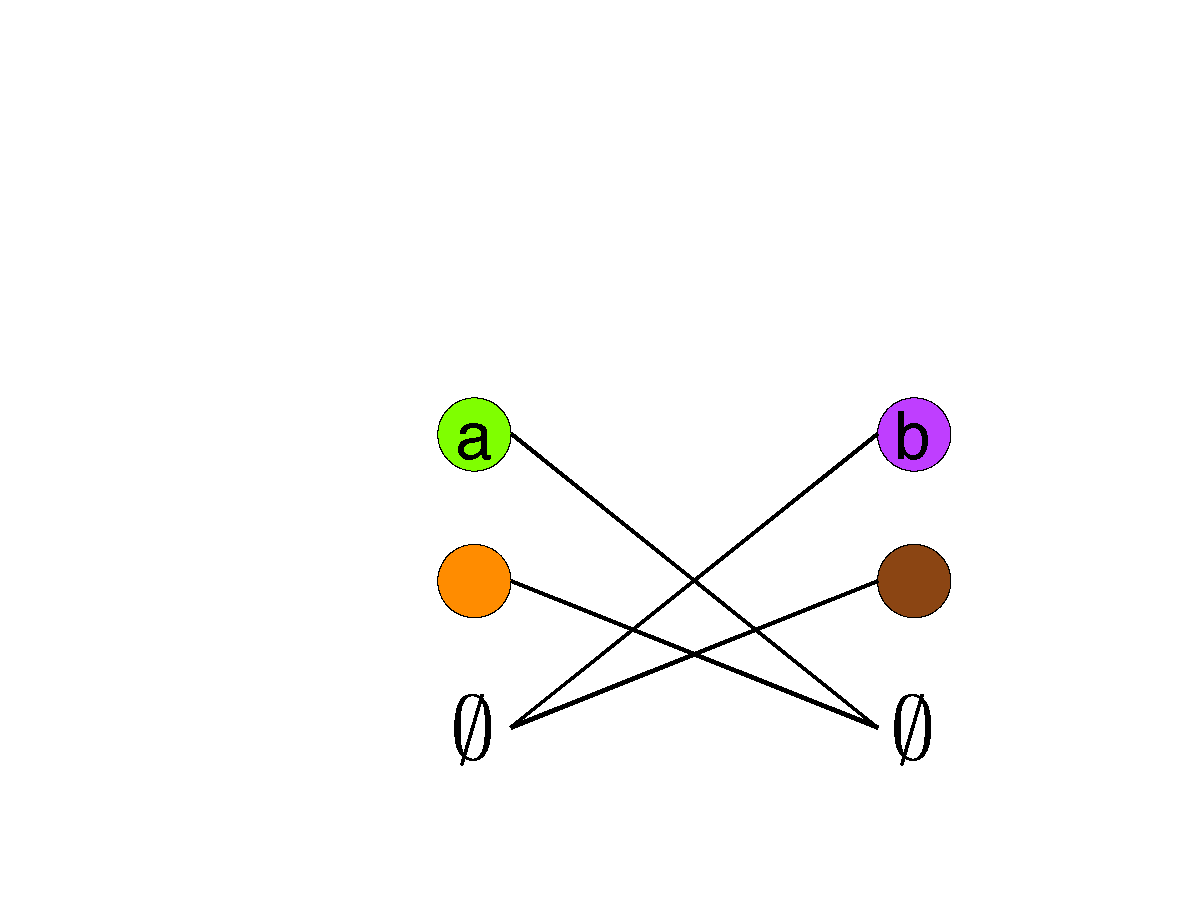}}
  \centerline{\footnotesize  \textbf{Hypothesis (iv) in cluster $C^{s_1}(H_3)$} }\medskip
\end{minipage}
\caption{
\footnotesize 
The hypotheses in cluster $C^{s_1}(H_3)$ are formed using the procedure discussed above.  The target arc shown
is contained in the $k=3$ best hypothesis.
  The clusters $C^{s_1}(H_1)$ and $C^{s_1}(H_3)$ exhaust all possible hypotheses in this problem.
 }
\label{fig:res2aa}
\end{figure}

Figures \ref{fig:res2a}-\ref{fig:res2aa} graphically illustrates cluster formation implied by arc ``$ab$'' in the association hypothesis shown at the top of the figure panel. Arc ``$ab$'' corresponds to $s_1$ in Fig. \ref{fig:illustrativeEg}.

\section{Cluster Probability Computations}
Here we show how to compute $Q_{ab}(C(H))$ given $H$, where $C(H)$ is the cluster containing $H$. We first consider the case where $H \in \HypsFix{ab}$, so $C(H) = C^{ab}_{H}$. We define the  \emph{switch value} $S(ab, xy)$ of two arcs as follows:

$$S(ab,xy) = \twopartdefother{\frac{L_{xb}L_{ay}}{L_{ab}L_{xy}}}{xb \in A \wedge ay \in A}{0}$$

Each hypothesis $H-ab-xy+xb+ay$ in $C^{ab}_{H}$  has likelihood $L(H)S(ab,xy)$. Thus, the total likelihood of the hypotheses in $C^{ab}_{H}$ is $L(H)\sum_{xy \in H }{S(ab,xy)}$. Since $C^{ab}_{H}$ contains only one hypothesis in $\HypsFix{ab}$ (namely H), we have $Q_{ab}(C(H)) = \frac{1}{\sum_{xy \in H }{S(ab,xy)}}$. (Since one of the terms in this denominator is always $S(ab,ab)$ which equals 1, so the denominator is always at least equal to 1.)

We now consider the case of $H \in \HypsHid{ab}$. Let $H^{\star} = H-aa^{\prime}-b^{\prime}b+b^{\prime}a^{\prime}+ab$, so $H \in C^{ab}_{H^{\star}}$ (recall 
$a^{\prime} \equiv adj_{\Hypk}(a)$;
see main text for definition of $adj_{\Hypk}(\cdot)$ ).

Define the function

$$S^{\star}(ab,xy) = \twopartdefother{0}{(a=x) \vee (b=y)}{S(ab,xy)}$$.

We also define the following functions:

$$S^{total}(ab,H) = \sum_{xy \in H }{S^{\star}(ab,xy)}$$

$$S^{extra}(ab,H) = \twopartdef{S^{\star}(ab,b^{\prime}a^{\prime})}{b^{\prime}a^{\prime} \in A}{\mathrm{undefined}}{b^{\prime}a^{\prime} \notin A}$$

Given these definitions, we can conclude that if ${b^{\prime}a^{\prime} \notin A}$, then $H^{\star}$ is undefined and $C_{ab}(H)=\{ H\}$ (therefore \newline $Q_{ab}\big(C(H)\big)=0$). Otherwise, we have:

$$Q_{ab}\big(C(H^{\star})\big) = \frac{1}{\sum_{(xy) \in H^{\star} }{S(ab,xy)}}$$
$$ = \frac{1}{\sum_{(xy) \in H-aa^{\prime}-b^{\prime}b+b^{\prime}a^{\prime}+ab }{S(ab,xy)}}$$
$$ = \frac{1}{1 + \sum_{(xy) \in H-aa^{\prime}-b^{\prime}b+b^{\prime}a^{\prime} }{S(ab,xy)}}$$
$$= \frac{1}{1 + \sum_{(xy) \in H+b^{\prime}a^{\prime} }{S^{\star}(ab,xy)}}$$
$$ = \frac{1}{1 + S^{extra}(ab,H) + \sum_{(xy) \in H }{S^{\star}(ab,xy)}}$$
$$ = \frac{1}{1 + S^{extra}(ab,H)+ S^{total}(ab,H) }$$.

Note that the above equation also applies to the case of $H \in \HypsFix{ab}$. If we define $H^{\star}$ as equal to $H$ for this case, we have $S^{extra}(ab,H) = 0$ and thus the equation above holds. 

\section{Complexity Analysis}

Suppose that we have a set of samples where each sample differs by at most $j$ changes (additions and deletions of arcs) from the previous sample. The amount of computation required to compute the cluster probabilities for a sample depends on whether it is the first sample (therefore we must perform all computations from scratch) or a subsequent sample (in this case, we can use information from the previous sample).

Additionally, the amount of computation required depends on how many target arcs we have. Since there is a different set of clusters for each target arc $ab$, we must keep track of $S^{total}(ab,H)$ and $S^{extra}(ab,H)$ for each target arc $ab$ separately. In many applications, we only have one or a small number of selected target arcs. In some applications, the target arcs are all the nonzero arcs in the best solution. In other applications, the target arcs are all the nonzero arcs in the problem.

We let $n$ be the maximum number of nodes in each frame and let $k$ be the maximum degree of the nodes.

Consider the case of the initial sample. There are at most $n$ nonzero arcs in the sample. For each such arc $xy$, the only arcs $ab$ such that $S^{\star}(ab,xy)$ can possibly be nonzero are those with $a$ adjacent to $y$ and $b$ adjacent to $x$. Since there are at most $k$ possible choices for $a$ and $k$ possible choices for $b$, we can make this computation in $O(k^{2})$ time, assuming that arcs can be accessed and their switch values computed in constant time. (Also, we can cache the computed list of $S^{\star}(ab,xy)$ values in case we change the arc $xy$ again in a later sample.) The computation of $S^{extra}(ab,H)$ can be done in constant time for each arc $ab$.

Thus, for the initial sample, we can compute the values of $S^{total}(ab,H)$ for all the arcs by computing the values of $S^{\star}(ab,xy)$ separately for each arc $xy \in H$. Since each arc $xy$ takes $O(k^{2})$ time, the entire computation takes  $O(nk^{2})$ time. Computation of $S^{extra}(ab,H)$ takes constant time for each arc, or a total of $O(nk)$ time, so the full computation still only takes $O(nk^{2})$ time. If one is concerned only with the arcs in the best solution, then the only arcs $ab$ that must be considered for any given $xy$ are the arcs $aa^{\prime}$ where $a$ is adjacent to $y$ and $a^{\prime}$ is the node that $a$ is associated with in the best solution. there are at most $k$ such arcs, so the computation can be done in $O(nk)$ time. Similarly, the computation of $S^{extra}(ab,H)$ adds only $O(n)$ time since there are at most $n$ arcs in the best solution, so the total time is still $O(nk)$. For computation of the association probability of a single arc $ab$, we just need to enumerate all the arcs $xy$ in the sample such that $S^{\star}(ab,xy) \ne 0$, and that can be done in $O(k)$ time.

For subsequent samples, when computing $S^{total}(ab,H)$, we can start with the previously computed totals and only need to update them by computing $S^{\star}(ab,xy)$ for the $xy$ that changed, and by assumption there are at most $j$ of those changes. As before, computing all the $S^{\star}(ab,xy)$ values for a single $xy$ takes $O(k^{2})$ time if considering all arcs as possible values for $ab$, takes $O(k)$ time if considering only arcs in the best solution as possible values for $ab$, and takes constant time if considering only a single target arc $ab$. Additionally, the number of target arcs $ab$ whose values of $S^{extra}(ab,H)$ have changed (and thus need to be recomputed) are $O(k^{2})$ for the all-arcs case, $O(k)$ for the arcs-in-best-solution case, and constant time for the single-arc case.
\begin{table}
\center
\begin{tabular}{ c | c | c}
  Target Arcs & First Sample Time & Subsequent Sample Time \\ \hline
  All Arcs & $O(nk^{2})$ & $O(jk^{2})$ \\
  Best Solution Arcs & $O(nk)$ & $O(jk)$ \\
  Single Arc & $O(k)$ & $O(j)$ \\
\end{tabular}
\caption{Time complexity with nonzero target arcs.}
\label{NonzeroComplexity}
\end{table}
 
 \section{Computational Details Specific to Applications Presented}
\subsection{Local MTT Association Problem Details}
To facilitate comparison with other computational approaches, we report the track and measurement ID numbers identifying the particles shown in Fig. 1 of the main text.
The $4 \times 3$ association problem  shown was formulated using  tracks IDs (193,194, 355,408) formed from frames 1-19 (formed using the tracking parameters indicated in the main text) with the particle measurement IDs (236,341,342) verified in frame 20.

\subsection{Stochastic Model}

Note that we simultaneously estimate the drift (induced by confinement) as well as measurement and thermal noise from a single candidate track (time spacing can be non-uniform).
Accounting for the above factors explains why our (unpruned)  estimated mean diffusion coefficient (0.0578 $\mu m^2/s$) is slightly lower than reported in Ref. \cite{Serge2008}.  Our estimation procedure is outlined below.   After tracks were formed we fit a stochastic differential equation (SDE) of the form:
  
\begin{align}
\label{eq:SDE}
dr_t= & \Phi F(r_t)dt+\sqrt{2}\sigma dB_t \\
\label{eq:SDEm}
\psi_{t_i}= & r_{t_i} +\epsilon_{t_i}  
\end{align}

\noindent   where the  two dimensional position of the particle measurements at time $t$ is denoted by the vector $r_t \equiv (x_t,y_t)^{\mathrm{T}}$. 
The subscript $t$ is used to index  time in the stochastic process.  In the SDE above, $B_t$ represents a standard 2D Brownian motion process. The notation $dB_t$   is used to denote an Ito \cite{raoDIFF}  stochastic integral (similarly for $dr_t$).   $F(r_t)$, is a vector representing the force associated with $r_t$, $\Phi$ is  a matrix quantifying effective friction, and $\sigma$ is related to the local diffusion coefficient \cite{pudi}.
 Stochastic effects inherent to SPT experiments, such as photon count uncertainty,  prevent us from directly observing $r_t$.  The only quantity we can directly measure  
in the lab is denoted by $\psi$.  This process is also indexed by time, e.g. , $\psi_{t_i} $, but we assume that measurements are only available at discrete time points.  The random variable $\epsilon_{t_i}  $ represents one 3D measurement noise realization at time $t_i$.

   We use the notation $\epsilon_{t_i}  \sim \mathcal{N}(0,\RR)$ to convey that $\epsilon_{t_i} $ is assumed to be distributed according to a mean zero 
   multivariate Gaussian  with covariance $\RR$.  
        Decoupling measurement noise from thermal noise inherent to all SPT trajectories is made possible by applying
   classic filtering  \cite{stengel1994}
   along with modern  SDE estimation \cite{Ait-Sahalia2010}.  These techniques ignore $dt$ terms (i.e., the drift).  Thermal and measurement noise estimates are then used to seed a Nelder-Mead algorithm (100 different initial guesses were used to help ensure that a local minimum was not encountered)
    searching for the global optimum parameter of the MLE (see \cite{raoDIFF} for cost function formulation). 

  The specific model we consider  is a linear SDE parameterized by a finite dimensional parameter vector denoted by $\color{red} \theta$.
   The  terms in
  Eqns. \ref{eq:SDE} and \ref{eq:SDEm}  are defined by the following  equations:

\begin{align}
  \label{eq:fluc}
  \Phi= &\sigma\sigma^{\mathrm{T}}/k_BT \\
\label{eq:SDEF}
  F(r)=&   \AA+\BB r   \\
\label{eq:SDEDiff}
 \sigma= & {\CC} \\
  \label{eq:SDEN}
  \ \epsilon \sim &  \mathcal{N}(0,\RR)
\end{align}

   Given the definition above, the  collection of parameters to be estimated by the data can be written as $\color{red} \theta$ $\equiv (\AA,\BB,\CC,\RR)$.   
  $\color{red} A$ is a 2D vector (i.e., $\color{red} A$ $ \in \mathbb{R}^2$). $\color{red} B$,  $\color{red} C$, and $\color{red} R$ are $2\times2$ real matrices. 
  The equations above were inspired by models in statistical physics \cite{SPA1,pudi}, e.g., $k_BT$ represents Boltzmann's 
 constant multiplied by the system temperature.
Also,   we  assume that the inertia of the particle can be ignored through a fluctuation dissipation relationship stated in Eqn. \ref{eq:fluc}, i.e., the tagged particle is in the ``overdamped" regime 
 \cite{pudi}.  
 Equation \ref{eq:SDEF} defines the parametric form of the effective local  linear force experienced by the overdamped particle and $\BB$ can be interpreted as an elasticity parameter.

\subsection{Gap-closing Parameters Utilized}

We used the default parameters in the publicly available code associated with Ref. \cite{Jaqaman2008}.  However, we increased the \url{gapCloseParam.timeWindow}$= 200$ in order liberally  allow for tracks to be stitched after quantum dot blinking (the UQ algorithm in conjunction with the $P_{GC}$ thresholding employed prevented overly aggressively associating track segments).  We also set \url{gapCloseParam.mergeSplit}$= 0$ to suppress merging and splitting;  this was only done to facilitate exposition, the tools introduced are readily applicable to scenarios including merge an split events in the gap-closing phase of track formation.

Since our method also assumes that the cost matrix connecting objects in two adjacent frames is the negative of a log likelihood, we transformed the squared distance variable $\delta^2_{IJ}$ (see Eqns. 3-4 in Ref. \cite{Jaqaman2008}) to $\delta_{IJ}$.  This transformation corresponds to the log likelihood of an exponential distribution with parameter one.  The choice of which probability distribution is subjective \cite{Jaqaman2008};  regardless of what distribution one desires to use for gap-closing, the correspondence method can be used to assess the uncertainty. 

\subsection{Spline fitting procedure}
Each track had a single diffusion coefficient fit to the track data.  The MLE estimate of the two-dimensional diffusion coefficient served as the ``dependent variable" and the time averaged $(x,y)$ position of each track served as the independent variables indexing the measurement (the individual tracks did not move far relative to inter-track variance).
The default thin plate smoothing parameters of the package in Ref. \cite{fields} was used to fit the resulting scatterplot surface (generalized cross validation smoothing was used).  \\

\newpage
 
\bibliographystyle{unsrt}
\bibliography{StatPapers,biomedtracking,tracking,running,ion_channels} 

\begin{thebibliography}{10}

\bibitem{Kim2006}
So~Yeon Kim, Zemer Gitai, Anika Kinkhabwala, Lucy Shapiro, and W~E Moerner.
\newblock {Single molecules of the bacterial actin MreB undergo directed
  treadmilling motion in Caulobacter crescentus.}
\newblock {\em Proceedings of the National Academy of Sciences of the United
  States of America}, 103(29):10929--34, July 2006.

\bibitem{Danuser2011}
Gaudenz Danuser.
\newblock {Computer vision in cell biology.}
\newblock {\em Cell}, 147(5):973--8, November 2011.

\bibitem{Meijering2012}
Erik Meijering, Oleh Dzyubachyk, and Ihor Smal.
\newblock {Methods for cell and particle tracking.}
\newblock {\em Methods in enzymology}, 504(February):183--200, January 2012.

\bibitem{Saxton2008}
Michael~J Saxton.
\newblock {Single-particle tracking: connecting the dots.}
\newblock {\em Nature Methods}, 5(8):671--2, August 2008.

\bibitem{Genovesio2006}
Auguste Genovesio, Tim Liedl, Valentina Emiliani, Wolfgang~J Parak, Mait\'{e}
  Coppey-Moisan, and Jean-Christophe Olivo-Marin.
\newblock {Multiple particle tracking in 3-D+t microscopy: method and
  application to the tracking of endocytosed quantum dots.}
\newblock {\em IEEE Transactions on Image Processing}, 15(5):1062--70, May
  2006.

\bibitem{Manley2008}
Suliana Manley, Jennifer~M Gillette, George~H Patterson, Hari Shroff, Harald~F
  Hess, Eric Betzig, and Jennifer Lippincott-Schwartz.
\newblock {High-density mapping of single-molecule trajectories with
  photoactivated localization microscopy.}
\newblock {\em Nature methods}, 5(2):155--7, February 2008.

\bibitem{Serge2008}
Arnauld Serg\'{e}, Nicolas Bertaux, Herv\'{e} Rigneault, and Didier Marguet.
\newblock {Dynamic multiple-target tracing to probe spatiotemporal cartography
  of cell membranes.}
\newblock {\em Nature Methods}, 5(8):687--94, August 2008.

\bibitem{Jaqaman2008}
Khuloud Jaqaman, Dinah Loerke, Marcel Mettlen, Hirotaka Kuwata, Sergio
  Grinstein, Sandra~L Schmid, and Gaudenz Danuser.
\newblock {Robust single-particle tracking in live-cell time-lapse sequences}.
\newblock {\em Nature Methods}, 5(8):695--702, 2008.

\bibitem{Saligrama2010}
Venkatesh Saligrama, Janusz Konrad, and Pierre-marc Jodoin.
\newblock {Video Anomaly Identification}.
\newblock {\em IEEE Signal Processing Magazine}, 27(5):18--33, September 2010.

\bibitem{Cezar2010}
Julio Cezar, Silveira Jacques, and Soraia~Raupp Musse.
\newblock {Crowd Analysis Using Computer Vision Techniques}.
\newblock {\em IEEE Signal Processing Magazine}, (September):66--77, 2010.

\bibitem{Fox2011}
Emily Fox, Erik~B. Sudderth, Michael~I. Jordan, and Alan~S. Willsky.
\newblock {Bayesian Nonparametric Inference of Switching Dynamic Linear
  Models}.
\newblock {\em IEEE Transactions on Signal Processing}, 59(4):1569--1585, April
  2011.

\bibitem{Kragel2012}
Bret Kragel, Shawn Herman, and Nick Roseveare.
\newblock {A Comparison of Methods for Estimating Track-to-Track Assignment
  Probabilities}.
\newblock {\em IEEE Transactions on Aerospace and Electronic Systems},
  48(3):1870--1888, July 2012.

\bibitem{Weigel2011}
Aubrey~V Weigel, Blair Simon, Michael~M Tamkun, and Diego Krapf.
\newblock {Ergodic and nonergodic processes coexist in the plasma membrane as
  observed by single-molecule tracking.}
\newblock {\em Proceedings of the National Academy of Sciences of the United
  States of America}, 108(16):6438--43, April 2011.

\bibitem{Deutsch2012}
Emily Deutsch, Aubrey~V Weigel, Elizabeth~J Akin, Phil Fox, Gentry Hansen,
  Christopher~J Haberkorn, Rob Loftus, Diego Krapf, and Michael~M Tamkun.
\newblock {Kv2.1 cell surface clusters are insertion platforms for ion channel
  delivery to the plasma membrane.}
\newblock {\em Molecular biology of the cell}, 23(15):2917--29, August 2012.

\bibitem{BP99}
S~Blackman and R~Popoli.
\newblock {\em {Design and Analysis of Modern Tracking Systems}}.
\newblock Artech House, Norwood, MA, 1999.

\bibitem{Reid1979}
D.~Reid.
\newblock {An algorithm for tracking multiple targets}.
\newblock {\em IEEE Transactions on Automatic Control}, 24(6):843--854,
  December 1979.

\bibitem{Poore1993}
Aubrey~B. Poore and Nenad Rijavec.
\newblock {A Lagrangian Relaxation Algorithm for Multidimensional Assignment
  Problems Arising from Multitarget Tracking}.
\newblock {\em SIAM Journal on Optimization}, 3(3):544--563, August 1993.

\bibitem{Poore2006}
A.B. Poore and S~Gadaleta.
\newblock {Some assignment problems arising from multiple target tracking}.
\newblock {\em Mathematical and Computer Modelling}, 43(9-10):1074--1091, May
  2006.

\bibitem{Blom_IMM_88}
H.A.P. Blom and Y.~Bar-Shalom.
\newblock The interacting multiple model algorithm for systems with {M}arkovian
  switching coefficients.
\newblock {\em Automatic Control, IEEE Transactions on}, 33(8):780 --783, 1988.

\bibitem{Godinez2009}
W~J Godinez, M~Lampe, S~W\"{o}rz, B~M\"{u}ller, R~Eils, and K~Rohr.
\newblock {Deterministic and probabilistic approaches for tracking virus
  particles in time-lapse fluorescence microscopy image sequences.}
\newblock {\em Medical Image Analysis}, 13(2):325--42, April 2009.

\bibitem{Chenouard2009}
Nicolas Chenouard, Alexandre Dufour, and Jean-Christophe Olivo-Marin.
\newblock {Tracking algorithms chase down pathogens.}
\newblock {\em Biotechnology journal}, 4(6):838--45, June 2009.

\bibitem{Chenouard2009a}
Nicolas Chenouard, I~Bloch, and Jean-Christophe Olivo-Marin.
\newblock {Multiple hypothesis tracking in microscopy images}.
\newblock In {\em IEEE International Symposium on Biomedical Imaging: From Nano
  to Macro. ISBI 2009}, pages 1346--1349, 2009.

\bibitem{Bertsekas1998}
Dimitri~P. Bertsekas.
\newblock {\em {Network Optimization: Continuous and Discrete Models
  (Optimization, Computation, and Control)}}.
\newblock Athena Scientific, 1998.

\bibitem{Mahler2003}
R.P.S. Mahler.
\newblock {Multitarget bayes filtering via first-order multitarget moments}.
\newblock {\em IEEE Transactions on Aerospace and Electronic Systems},
  39(4):1152--1178, October 2003.

\bibitem{Panta2004}
Kusha Panta, Ba-Ngu Vo, Sumeetpal Singh, and Arnaud Doucet.
\newblock {<title>Probability hypothesis density filter versus multiple
  hypothesis tracking</title>}.
\newblock pages 284--295, August 2004.

\bibitem{JValg}
R.~Jonker and A.~Volgenant.
\newblock A shortest augmenting path algorithm for dense and sparse linear
  assignment problems.
\newblock {\em Computing}, 38(4):325--340, November 1987.

\bibitem{drummond90}
{O.E.} Drummond, {D.A.} Castanon, and {M.S.} Bellovin.
\newblock Comparison of {2D} assignment algorithms for sparse, rectangular,
  floating point, cost matrices.
\newblock {\em J. SDI Panels on Tracking}, 4:81--97, 1990.

\bibitem{Danchick2006}
R.~Danchick and G.E. Newnam.
\newblock {Reformulating Reid's MHT method with generalised Murty K-best ranked
  linear assignment algorithm}.
\newblock {\em IEE Proceedings - Radar, Sonar and Navigation}, 153(1):13,
  February 2006.

\bibitem{billingsley}
P.~Billingsley.
\newblock {\em Convergence of probability measures}.
\newblock Wiley, 1968.

\bibitem{SPAdsDNA}
Christopher~P. Calderon, {W.-H.} Chen, N.~Harris, {K.J.} Lin, and {C.-H.}
  Kiang.
\newblock Quantifying {DNA} melting transitions using single-molecule force
  spectroscopy.
\newblock {\em J. Phys.: Condens. Matter}, 21:034114, 2009.

\bibitem{SPAfric}
Christopher~P. Calderon, N.~Harris, {C.-H.} Kiang, and {D.D.} Cox.
\newblock Analyzing single-molecule manipulation experiments.
\newblock {\em J. Mol. Recognit.}, 22:356, 2009.

\bibitem{pudi}
Christopher~P. Calderon, {J.G.} Martinez, {R.J.} Carroll, and {D.C.} Sorensen.
\newblock P-splines using derivative information.
\newblock {\em Multiscale Model. Simul.}, 8:1562--1580, 2010.

\bibitem{Arulampalam2002}
M.S. Arulampalam, S.~Maskell, N.~Gordon, and T.~Clapp.
\newblock {A tutorial on particle filters for online nonlinear/non-Gaussian
  Bayesian tracking}.
\newblock {\em IEEE Transactions on Signal Processing}, 50(2):174--188, 2002.

\bibitem{Tamkun2007}
Michael~M Tamkun, Kristen M~S O'connell, and Annah~S Rolig.
\newblock {A cytoskeletal-based perimeter fence selectively corrals a
  sub-population of cell surface Kv2.1 channels.}
\newblock {\em J. Cell Science}, 120(Pt 14):2413--23, July 2007.

\bibitem{Weigel2012}
Aubrey~V. Weigel, Shankarachary Ragi, Michael~L. Reid, Edwin K.~P. Chong,
  Michael~M. Tamkun, and Diego Krapf.
\newblock {Obstructed diffusion propagator analysis for single-particle
  tracking}.
\newblock {\em Physical Review E}, 85(4):041924, April 2012.

\bibitem{SPAfilter}
Christopher~P. Calderon, N.~Harris, {C.-H.} Kiang, and {D.D.} Cox.
\newblock Quantifying multiscale noise sources in single-molecule time series.
\newblock {\em J. Phys. Chem. B}, 113:138, 2009.

\bibitem{SPAgof}
Christopher~P. Calderon.
\newblock {Detection of Subtle Dynamical Changes Induced by Unresolved
  Conformational Coordinates in Single-Molecule Trajectories via
  Goodness-of-Fit Tests}.
\newblock {\em J Phys Chem B}, 114:3242--3253, 2010.

\bibitem{raoDIFF}
{B.L.S.} Prakasa~Rao.
\newblock {\em Statistical Inference for Diffusion Type Processes}.
\newblock Arnold Publishers, London, 1999.

\bibitem{stengel1994}
R.F. Stengel.
\newblock {\em Optimal control and estimation}.
\newblock Dover Publications, Toronto, Ontario, 1994.

\bibitem{Ait-Sahalia2010}
Yacine A\"{\i}t-Sahalia, Jianquing Fan, and Dacheng Xiu.
\newblock {High-Frequency Covariance Estimates With Noisy and Asynchronous
  Financial Data}.
\newblock {\em Journal of the American Statistical Association},
  105(492):1504--1517, December 2010.

\bibitem{SPA1}
Christopher~P. Calderon.
\newblock On the use of local diffusion for path ensemble averaging in
  potential of mean force computations.
\newblock {\em J. Chem. Phys.}, 126:084106, 2007.

\bibitem{fields}
{Fields Development Team (2006)}.
\newblock {\em fields: Tools for Spatial Data}.
\newblock \url{http://www.cgd.ucar.edu/Software/Fields}, National Center for
  Atmospheric Research, Boulder, CO., 2006.

\end{thebibliography}

\end{document}